\documentclass[reprint,secnumarabic,amsmath, amssymb, nofootinbib, aps, pre]{revtex4-1}

\usepackage{graphicx}
\usepackage{lipsum}
\usepackage{xcolor}

\DeclareMathOperator{\diag}{diag}

\DeclareMathOperator{\sign}{sign}

\usepackage{ifthen}
\newcounter{currentref}

\makeatletter
\newcommand{\refgen}[3][]{%
    \@ifundefined{r@eq:#2}{
        #3(\ref{eq:#2})
    }{%
        \ifthenelse{\equal{#1}{}}{
            #3(\ref{eq:#2})%
        }{
            \setcounter{currentref}{\ref{eq:#2}}%
            \addtocounter{currentref}{#1}%
            #3(\ref{eq:#2})-(\arabic{currentref})%
        }%
    }%
}
\makeatother

\newcommand{\refeq}[2][]{\refgen[#1]{#2}{eq.~}}
\newcommand{\refno}[2][]{\refgen[#1]{#2}{}}

\newcommand{\er}{\mathbb{R}}    
\newcommand{\es}{\{-1,1\}}      

\renewcommand{\vec}[1]{\mathbf{#1}}
\newcommand{\mat}[1]{\mathbf{#1}}

\newcommand{\bh}{\vec{h}}   
\newcommand{\bJ}{\mat{J}}   
\newcommand{\bs}{\vec{s}}   
\newcommand{\br}{\vec{r}}   
\newcommand{\ES}{\boldsymbol{\rm m}}        
\newcommand{\CovS}{\boldsymbol{\rm C}}      
\newcommand{\ER}{\boldsymbol{\mu}}          
\newcommand{\CovR}{\boldsymbol{\Sigma}}     

\newcommand{\btheta}{\boldsymbol{\theta}}        

\newcommand{\mI}{\mat{I}}                   
\newcommand{\mD}{\mat{D}}                   

\newcommand{\Bern}{\mathcal{B}}     
\newcommand{\Norm}{\mathcal{N}}     
\newcommand{\ZI}{Z_I}    
\newcommand{\ZB}{Z_B}    

\newcommand{\dbr}{{\rm d}\br}

\newcommand{\dx}{{\rm d}x}

\newcommand{\pE}{\mathrm{E}}        
\newcommand{\Var}{\mathrm{Var}}
\newcommand{\Cov}{\mathrm{Cov}}

\newcommand{\tkappa}{\kappa}
\newcommand{\tlambda}{\lambda}
\newcommand{\tnu}{\nu}

\newcommand{\cavi}{_{\backslash i}}
\newcommand{\hcav}{\tilde{r}}
\newcommand{\tcavij}{\theta_i^{(j)}}
\newcommand{\tcavji}{\theta_j^{(i)}}

\usepackage{amsthm}
\theoremstyle{plain}

\newcommand{\rev}[1]{#1}	

\begin{document}

\title{Ising distribution as a latent variable model}%

\begin{abstract}

During the past decades, the Ising distribution has attracted interest in many applied disciplines, as the maximum entropy distribution associated to any set of correlated binary (`spin') variables with observed means and covariances. However, numerically speaking, the Ising distribution is unpractical, 
so alternative models are often preferred to handle correlated binary data.
One popular alternative, especially in life sciences, is the \emph{Cox distribution} (or the closely related \emph{dichotomized Gaussian distribution} and \emph{log-normal Cox point process}), where the spins are generated independently conditioned on the drawing of a latent variable with a multivariate normal distribution.
This article explores the conditions for a principled replacement of the Ising distribution by a Cox distribution.
It shows that the Ising distribution itself can be treated as a latent variable model, and it explores when this latent variable has a quasi-normal distribution. A variational approach to this question reveals a formal link with classic mean field methods, especially Opper and Winther's \emph{adaptive TAP} approximation. This link is confirmed by weak coupling (Plefka) expansions of the different approximations, and then by numerical tests. 
Overall, this study suggests that an Ising distribution can be replaced by a Cox distribution in practical applications, precisely when its parameters lie in the `mean field domain'.
\end{abstract}

\author{Adrien Wohrer}%
\email{adrien.wohrer@uca.fr\\Accepted for publication in {\it Physical Review E}}
\affiliation{Universit\'e Clermont Auvergne, CNRS, SIGMA Clermont, Institut Pascal, F-63000 Clermont-Ferrand, France.} 

\maketitle

\section{Introduction}

During the last decades, the Ising distribution has been used in several disciplines such as statistics (under the name \emph{quadratic exponential model}) \cite{Cox1972,ZhaoPrentice1990}, machine learning (under the name \emph{Boltzmann machine}) \cite{Ackley1987}, information processing \cite{Mezard1987,OpperSaad2001,Nishimori2001}, biology \cite{Weigt2009} and neurosciences, where
it has been proposed as a natural model for the 
spike-based activities of interconnected neural populations \cite{Schneidman2006,Ohiorhenuan2010}.
In most of these applications, classic assumptions from statistical physics do not hold (e.g., arrangement on a rectangular lattice, uniform couplings, independently distributed couplings, zero external fields, etc.), and even old problems have to be revisited, such as efficiently simulating the Ising distribution \cite{Hukushima1996,WangLandau2001} or inferring its parameters from data \cite{KappenRodriguez1998,Sessak2009,Roudi2009,Cocco2011}.

In this article I will consider the Ising probability distribution over a set of spins $\bs=(s_1,\dots,s_N)\in\es^N$ defined as
\begin{equation}
\label{eq:ising1}
P(\bs|\bh,\bJ)=\frac{1}{\ZI}\exp\Big(\sum_{i=1}^N h_is_i+\frac{1}{2}\sum_{i,j=1}^N J_{ij}s_is_j\Big),
\end{equation}
with parameters $\bh\in\er^N$ (external fields) and $\bJ$ an $N\times N$ symmetric matrix (coupling weights), $\ZI(\bh,\bJ)$ being the corresponding partition function. In this formulation, diagonal elements $J_{ii}$ can be nonzero without influencing the distribution, simply adding a constant term $\sum_i J_{ii}/2$ 
to both the exponent and $\log(Z_I)$.

I will note $(\ES,\CovS)$ the two first centered moments of the distribution, that is, for all indices $i,j$,
\begin{equation*}
m_i=\pE(s_i) \quad,\quad C_{ij}=\pE(s_is_j)-\pE(s_i)\pE(s_j).
\end{equation*}
The essential interest of the Ising distribution, in all disciplines mentioned above, is its \emph{maximum entropy} property~: whenever a dataset of $N$ binary variables has measured moments $(\ES,\CovS)$, a single distribution of the form (\ref{eq:ising1}) is guaranteed to exist which matches these moments, and furthermore it has maximal entropy under this constraint.

\rev{%
Unfortunately, the Ising distribution is numerically unwieldy. 
The simple act of drawing samples from the distribution already requires to set up lengthy Markov Chain Monte Carlo (MCMC) schemes.
Besides, there is no simple analytical link between parameters $(\bh,\bJ)$ and resulting moments $(\ES,\CovS)$. Given natural parameters $(\bh,\bJ)$, the \emph{direct Ising problem} of estimating $(\ES,\CovS)$ can only be solved by numerical sampling from MCMC chains. Given $(\ES,\CovS)$, the \emph{inverse Ising problem} of retrieving $(\bh,\bJ)$ can only be solved by gradient descent based on numerous iterations of the direct problem, a procedure known as \emph{Boltzmann learning}. In practice, this means that the Ising distribution cannot be parametrized easily from observed data.
}


For this reason, in spite of the Ising model's theoretical attractiveness when dealing with binary variables, alternative models are generally preferred, which are numerically more convenient.
In one such family of alternative models,
$N$ \emph{latent variables} $\br=(r_1,\dots,r_N)\in\er^N$ are drawn from a multivariate normal distribution 
\begin{equation*}
\Norm(\br|\ER,\CovR)=|2\pi\CovR|^{-1/2}\exp\Big(-\frac{1}{2}(\br-\ER)^\top\CovR^{-1}(\br-\ER)\Big)
\end{equation*}
and then used to generate $N$ spins independently. The simplest option, setting $s_i=\sign(r_i)$ deterministically, yields the \emph{dichotomized Gaussian} distribution \cite{Pearson1909,CoxWermuth2002}, which has enjoyed recent popularity as a replacement for the Ising distribution when modeling neural spike trains, as it is easy to sample and to parametrize from data \cite{Amari2003,Macke2011}.

Slightly more generally, each variable $r_i$ can serve as an intensity to draw the corresponding spin $s_i$ following a Bernoulli distribution~:
\begin{equation*}
\Bern(\bs|\br)=\frac{1}{\ZB}\exp\Big(\sum_{i=1}^N r_is_i\Big)
\end{equation*}
with partition function
\begin{equation*}
\ZB(\br)=\prod_i\big(e^{r_i}+e^{-r_i}\big).
\end{equation*}
In statistics, this is the model underlying logistic regression, as introduced by Cox \cite{Cox1958}, so I will refer to it as
the \emph{Cox distribution}~:
\begin{align}
Q(\br|\ER,\CovR) &= \Norm(\br|\ER,\CovR), \label{eq:cox1}\\
Q(\bs|\br) &= \Bern(\bs|\br), \label{eq:cox2}
\end{align}
 with $\ER$ any vector in $\er^N$ and $\CovR$ any $N\times N$ symmetric definite positive matrix.
Note that the dichotomized Gaussian corresponds to a limiting case of the Cox distribution,
when the scaling of variables $r_i$ tends to $+\infty$
\footnote{%
The dichotomized Gaussian of parameters $(\ER,\CovR)$ is the limit of the Cox distribution of parameters $(\lambda\ER,\lambda^2\CovR)$ when $\lambda\rightarrow+\infty$. Conversely, note that a Bernoulli variable $S\sim\mathcal{B}(r)$ can be generated as $S={\rm sign}(r+X)$ where $X$ follows the \emph{logistic distribution} of density function $f(x)=\frac{1}{2}\cosh^{-2}(x)$, which resembles closely the normal distribution $\mathcal{N}(x|0,\kappa^2)$ with $\kappa\simeq 0.85$. As a result, any Cox distribution of parameters $(\ER,\CovR)$ is decently approximated by a dichotomized Gaussian of parameters $(\ER,\CovR+\kappa^2\mI)$.}.

Both the Ising (eq.~(\ref{eq:ising1})) and Cox (eq.~(\ref{eq:cox1})-(\ref{eq:cox2})) distributions can be generalized to point processes, by taking a suitable limit when the $N$ indexed variables tend to a continuum \cite{DaleyVereJones2003}. These are respectively known as the \emph{Gibbs process} \cite{DaleyVereJones2003} and \emph{log Gaussian Cox process} \cite{Cox1955,Moller1998}. The latter, much simpler to handle in practice, is used in various applied fields such as epidemiology, geostatistics \cite{Diggle2013} and neurosciences, to model neural spike trains \cite{KruminShoham2009,Brette2009}.

\rev{
To summarize, in practical application, Cox models (including the dichotomized Gaussian, and Cox point processes) are often preferred to the corresponding maximum entropy distributions (Ising distribution, Gibbs point process) because they are easier to sample, and to parametrize from a set of observed data. 
However, to date, we have little analytical insights into the link between the two families of distributions. For example, given some dataset, we cannot tell in advance how similar the Cox and Ising models fitting this data would be. The goal of this article is to investigate this link.

I first show that the Ising distribution itself can be viewed as a latent variable model, which differs from a Cox distribution only because of the non-Gaussian distribution of its latent variable (Section \ref{sec:latent}). This allows to derive simple relations between an Ising distribution and its `best-fitting' Cox distributions, in two possible senses (Section \ref{sec:cox}).
In particular, the variational approach for targeting a best-fitting Cox distribution displays formal similarities with classic mean-field methods which aim at approximating the Ising moments $(\ES,\CovS)$ (Section \ref{sec:meanfield}). Numerical simulations reveal that both types of approximations, despite their seemingly different goals, are efficient in roughly the same domain of parameters (Section \ref{sec:tests}).
Thus, an Ising distribution can be replaced in practical applications by a Cox distribution, precisely if its parameters lie in the `mean field domain'.
}

\section{The Ising latent field}
\label{sec:latent}

Given any vector $\bh\in\er^N$ and $N\times N$ symmetric, \emph{definite positive} matrix $\bJ$, we will consider the following probability distribution over $\br\in\er^N$ and $\bs\in\es^N$~:
\begin{equation}
\label{eq:IL}
P(\bs,\br)=\frac{1}{Z}\exp\Big(-\frac{1}{2}(\br-\bh)^\top\bJ^{-1}(\br-\bh)+\br^\top\bs\Big),
\end{equation}
with $Z$ ensuring proper normalization.

Marginalizing out variable $\bs$ yields
\begin{align}
P(\br) &=\frac{\ZB(\br)}{Z}\exp\Big(-\frac{1}{2}(\br-\bh)^\top\bJ^{-1}(\br-\bh)\Big), \label{eq:latent}\\
P(\bs|\br) &= \Bern(\bs|\br).            \label{eq:ScondR}
\end{align}
Conversely, completing the square in eq.~(\ref{eq:IL}) and marginalizing out variable $\br$ yields
\begin{align}
P(\bs) &= \frac{|2\pi\bJ|^{1/2}}{Z}\exp\Big(\bh^\top\bs+\frac{1}{2}\bs^\top\bJ\bs\Big), \label{eq:ising}\\
P(\br|\bs) &= \Norm(\br|\bh+\bJ\bs,\bJ). \label{eq:RcondS}
\end{align}
From eq.~(\ref{eq:ising}), the resulting spins $\bs$ are distributed according to the Ising distribution of parameters $(\bh,\bJ)$.

The introduction of field variables $r_i$ has long been known in statistical physics, as a mathematical construct to express the Ising partition function $Z_I$ in an integral form \cite{OpperSaad2001}. Indeed, equating the respective expressions for $Z$ imposed by eq.~(\ref{eq:latent}) and (\ref{eq:ising}), we obtain the elegant formula
\begin{equation}
\label{eq:ZIint}
\ZI(\bh,\bJ) = \int_{\br\in\er^N}\ZB(\br)\Norm(\br|\bh,\bJ)\dbr
\end{equation}
expressing $\ZI$ as the convolution of $\ZB$ with a Gaussian kernel of covariance $\bJ$.
This formula can be used as a justification of classic mean field equations \cite{OpperSaad2001}, and more generally to derive the diagrammatic (i.e., Taylor) expansion of $\ZI$ as a function of $\bJ$ \cite{VasilevRadzhabov1974}.

In this work instead, I view the $r_i$ as a set of probabilistic variables in their own right, coupled to the Ising spin variables, through eq.~(\ref{eq:latent})-(\ref{eq:RcondS}).
Given a spin configuration $\bs$, variable $\br$ is normally distributed (eq.~(\ref{eq:RcondS})). Thus, the overall distribution $P(\br)$ is a mixture of Gaussians with $2^N$ components, where the component associated to spin configuration $\bs$ has weight $P(\bs)$. More compactly, $P(\br)$ can be expressed with eq.~(\ref{eq:latent}).

Given some configuration $\br$, the spins $\bs$ can simply be drawn independently following a Bernoulli distribution (eq.~(\ref{eq:ScondR})), so the Ising distribution $P(\bs)$ itself can be viewed as a latent variable model, based on hidden variables $r_i$. It departs from a Cox distribution (eq.~(\ref{eq:cox1})-(\ref{eq:cox2})) only through the fact that the fields' distribution $P(\br)$ is not normal, in general.

\section{Cox approximations to the Ising distribution}
\label{sec:cox}


This article investigates the possible replacement of the Ising distribution by a Cox distribution. 
With the above reformulation, this amounts to approximating the Ising latent field distribution $P(\br)$ by a well-chosen multivariate normal $Q(\br)=\mathcal{N}(\br|\ER,\CovR)$. I will now discuss two possible choices in this regard. 

From here on, I will note $(\ES,\CovS)$ for (any approximation of) the first moments of a spin variable $\bs$, and $(\ER,\CovR)$ for (any approximation of) the first moments of a field variable $\br$. I will distinguish the true moments of the Ising distribution $P(\bs)$ with a star~: $(\ES^\star,\CovS^\star)$. Likewise, the true moments of the corresponding Ising latent field $P(\br)$ follow, from \refeq{RcondS}~:
\begin{align}
\ER^\star &= \bh+\bJ\ES^\star, \label{eq:ERSstar} \\
\CovR^\star  &= \bJ+\bJ\CovS^\star\bJ. \label{eq:CovRSstar}
\end{align}

\subsubsection*{Optimal Cox distribution}

Arguably, the optimal approximation of $P(\br)$ by a normal distribution $Q(\br)=\Norm(\ER,\CovR)$ is achieved by equating their moments, i.e., setting $(\ER,\CovR)=(\ER^\star,\CovR^\star)$. I will refer to this choice as the \emph{optimal} Cox distribution.
Its qualification as `optimal' stems from the observation that the following KL divergence
\begin{equation}
{\rm KL}_{\br}(P||Q) = \int_{\br\in\er^N} P(\br)\ln\frac{P(\br)}{Q(\br|\ER,\CovR)}{\rm d}\br
\end{equation}
is minimized when $(\ER,\CovR)=(\ER^\star,\CovR^\star)$. Thus, in terms of information geometry, the resulting distribution $Q(\br|\ER^\star,\CovR^\star)$ 
is the nearest neighbor of the latent Ising distribution $P(\br)$ in the family of normal distributions.

Unfortunately, the optimal Cox distribution is unpractical to characterize, as $(\ES^\star,\CovS^\star)$ can only be estimated by lengthy Monte-Carlo simulation.
In the scope of this article, its study will only be of theoretical interest~: it allows to quantify the intrinsic effect of assuming a Gaussian shape for $P(\br)$.

\subsubsection*{Variational Cox approximation}

More practically, one may require to approximate
an Ising distribution $P(\bs)$ by a Cox distribution, assuming only knowledge of its natural parameters $(\bh,\bJ)$.
The \emph{variational} approach to this problem consists in choosing $(\ER,\CovR)$ that minimize the reversed Kullback-Leibler divergence
\begin{equation}
\label{eq:KLQP}
{\rm KL}_{\br}(Q||P) = \int_{\br\in\er^N} Q(\br|\ER,\CovR)\ln\frac{Q(\br|\ER,\CovR)}{P(\br)}{\rm d}\br.
\end{equation}
With some straightforward algebra, one can establish the derivatives of ${\rm KL}_{\br}(Q||P)$ with respect to $\ER$ and $\CovR$, and thus its stationary points $(\ER,\CovR)$.

Given the fundamental relation \refno{ERSstar}-\refno{CovRSstar} between spin and field moments in the Ising distribution, it is natural to reparametrize the Cox parameters $(\ER,\CovR)$ by the `spin moment' parameters $(\ES,\CovS)$ such that
\begin{align}
\ER &= \bh+\bJ\ES, \label{eq:ERS} \\
\CovR  &= \bJ+\bJ\CovS\bJ. \label{eq:CovRS}
\end{align}
Then, the values of $(\ES,\CovS)$ at the stationary points of ${\rm KL}_{\br}(Q||P)$ are characterized by the following, fixed point equation~:
\begin{align}
m_i &= \int_{x\in\er} \tanh\left(\mu_i+x\sqrt{\Sigma_{ii}}\right)\phi(x)\dx, \label{eq:ES_var}\\
d_i &= \int_{x\in\er} \left(1-\tanh^2\left(\mu_i+x\sqrt{\Sigma_{ii}}\right)\right)\phi(x)\dx, \label{eq:di}\\
(\CovS^{-1})_{ij} &= d_i^{-1}\delta_{ij} -J_{ij},\label{eq:CovS_var}
\end{align}
\rev{with $\phi(x)=\Norm(x|0,1)$ the standard one-dimensional normal distribution.}

Equations \refno{ERS}-\refno{CovS_var} can be solved by an iterative fixed point method on variables $\{m_i,d_i\}_{i=1\dots N}$ (see Appendix \ref{app:numerical}). At the solution, the Cox distribution $Q(\ER,\CovR)$ provides an approximation to the Ising distribution $P(\bh,\bJ)$.

The formulas \refno{ES_var}-\refno{di} are conceptually simple~: $m_i$ (resp. $d_i$) is obtained as the average of $\tanh(r)$ (resp. $1-\tanh^2(r)$) using a Gaussian kernel, centered around $r=\mu_i$ with variance $\Sigma_{ii}$. Their estimation at any required precision is straightforward, using numerical integration. However, when repeated computations are required, it is faster to use approximate formulas, given in Appendix \ref{sec:tanherf}.

Figure \ref{fig2D} illustrates the nature of the latent field $P(\br)$, and of its Cox approximations, on a 2-spin toy model. The optimal Cox distribution is unique by construction, but the variational Cox approximation can have multiple solutions (panel b), a classic feature of variational methods based on minimizing reversed KL divergence \cite{Bishop2006}.

\begin{figure}
\includegraphics[width=\linewidth]{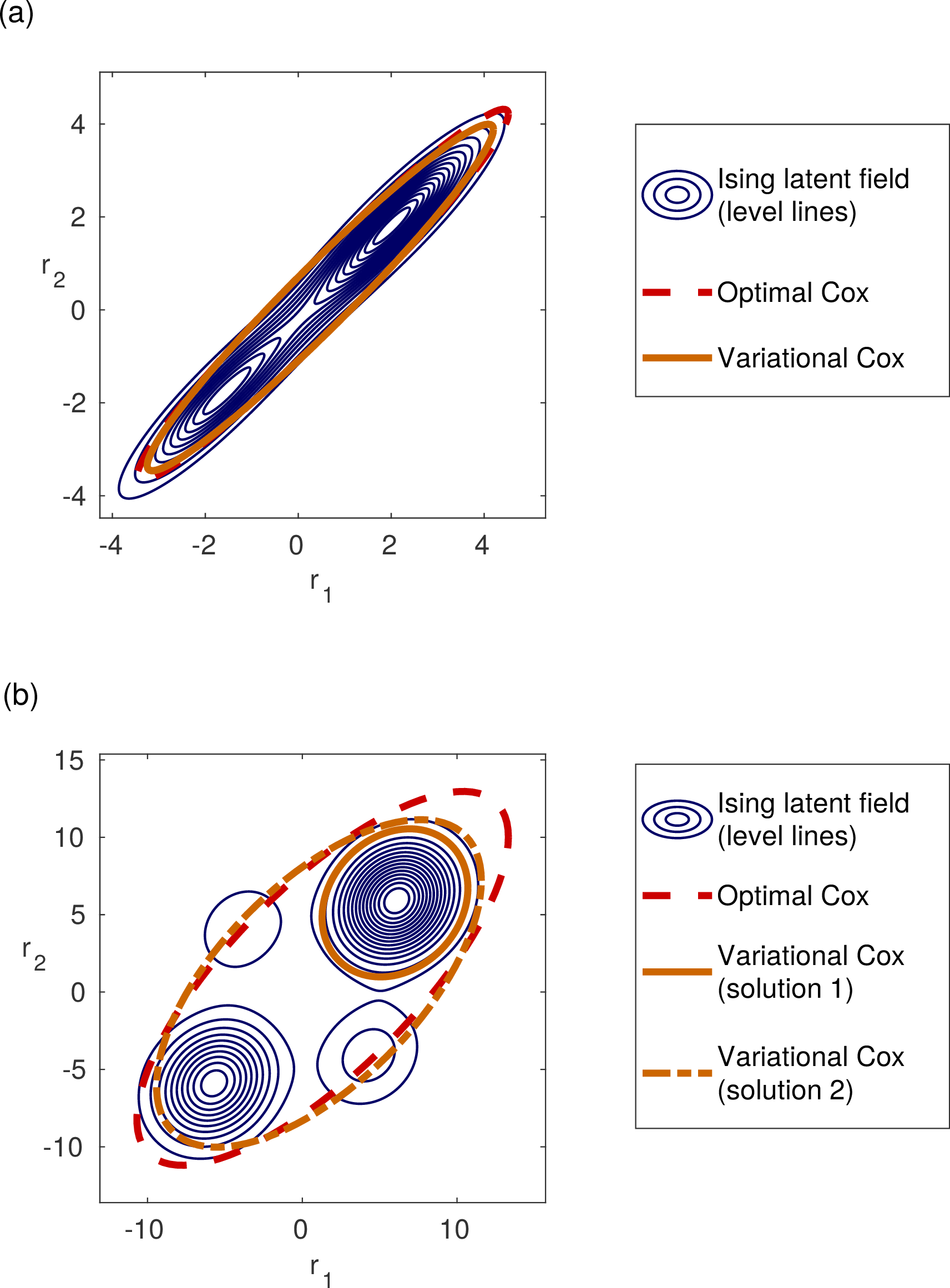}
\caption{\label{fig2D}
Ising latent field and its normal approximations on a toy model with $N=2$. Normal distributions are materialized by their 2-$\sigma$ level line.
Classic Ising parameters $(h_1,h_2,J_{12})=(0.2,0,0.9)$. In (a), diagonal couplings are fixed at $J_{11}=J_{22}=1$, whereas in (b) they are fixed at $J_{11}=J_{22}=5$.
The Ising spin distribution $P(\bs)$ is the same in both cases, but not the field distribution $P(\br)$ and subsequent Cox approximations. In (b), the variational Cox equation has multiple solutions -- two of which are displayed.
}
\end{figure}

\subsubsection*{Choice of $\diag(\bJ)$}

\rev{The framework developed above requires strictly positive diagonal coupling values $J_{ii}$, large enough to ensure that matrix $\bJ$ is definite positive.
While these diagonal values do not influence the Ising distribution $P(\bs)$ over spins, a different choice of $\diag(\bJ)$ leads to a different latent field distribution $P(\br)$, and thus to a different Cox approximation (see Figure \ref{fig2D}).
This naturally raises the question of what self-couplings $J_{ii}$ represent in the latent field formalism, and how they should be chosen in practice. Here, I only detail one concrete proposal for this choice, and defer more general considerations to the Discussion.

In the perspective of this work, the choice should be made to optimize the resemblance of $P(\br)$ with a normal distribution.}
From eq.~(\ref{eq:RcondS}), larger values of $\bJ$ increase the overall separation of the $2^N$ components in  $P(\br)$ and thus, its divergence from a normal distribution.
In the extreme case where $\diag(\bJ)$ is very large, the $2^N$ components of $P(\br)$ display no overlap at all~: see Figure \ref{fig2D}(b).

Consequently, the general prescription is that $\diag(\bJ)$ should be kept as small as possible. Given a fixed set of off-diagonal weights $\{J_{ij}\}_{i<j}$, a principled approach is to choose $\diag(\bJ)$ with the following procedure~:
\begin{equation}
\label{eq:TrJ}
\underset{\{J_{ii}\}}{\text{minimize}}
\quad \sum_i J_{ii} \quad \text{subject to} \quad
 \bJ \succeq 0.
\end{equation}
This is a well known convex problem, which can be solved efficiently \cite{Boyd2004}.
Afterwards, a small ridge term $\lambda\mI$ can be added to $\bJ$ 
to make it \emph{strictly} definite positive.

\subsubsection*{Moments of the Cox distribution}

A note of caution is also required on the interpretation of variables $(\ES,\CovS)$ in the variational Cox approximation, \refeq{ERS}-\refno{CovS_var}.
By eq.~(\ref{eq:cox1})-(\ref{eq:cox2}), the first spin moment of Cox distribution $Q(\bs|\ER,\CovR)$ is
\begin{align}
\pE_Q(s_i) &= \int_{r\in\er} \tanh(r)\Norm(r|\mu_i,\Sigma_{ii}){\rm d}r, \label{eq:EScox}
\end{align}
and we recognize \refeq{ES_var}. Thus, at the fixed point of \refeq{ERS}-\refno{CovS_var}, $m_i$ corresponds to the first moment of the approximating Cox distribution.

In contrast, at the fixed point of \refeq{ERS}-\refno{CovS_var}, $\CovS$ is \emph{not} the spin covariance of the Cox distribution $Q(\bs|\ER,\CovR)$. That would be computed (for $i\neq j$) as
\begin{align}
\Cov_Q(s_i,s_j) &= \iint_{(r,t)\in\er^2} \tanh(r)\tanh(t)\Norm(r,t|\ER^{(ij)},\CovR^{(ij)}){\rm d}r{\rm d}t \nonumber\\
& \quad - \pE_Q(s_i)\pE_Q(s_j),\label{eq:CovScox}
\end{align}
with $\ER^{(ij)},\CovR^{(ij)}$ the two-dimensional restrictions of $\ER,\CovR$ at indices $(i,j)$.

Thus, for given Cox parameters $(\ER,\CovR)$, there are two possible predictions for the Ising covariance matrix $\CovS$~: the `forward' prediction of \refeq{CovScox} (spin covariance matrix in the Cox distribution) and the `backward' prediction of \refeq{CovRS} (spin covariance in the Ising model which would give rise to field covariance matrix $\CovR$).
If $Q(\br|\ER,\CovR)$ is a good approximation of $P(\br)$, we expect both predictions to be very close.
And indeed, the discrepancy between the two predictions of $\CovS$ is a good indicator of whether the approximation was successful (see Supplementary Material).

\section{Comparison with mean field approximations}
\label{sec:meanfield}

In the variational Cox approximation, \refeq{ERS}-\refno{CovS_var}, variables $(\ES,\CovS)$ constitute an approximation for the moments of the Ising distribution $P(\bh,\bJ)$. The goodness of fit wrt. exact Ising moments $(\ES^\star,\CovS^\star)$ constitutes a simple measure of how well distribution $P$ is approximated by the Cox distribution $Q(\ER,\CovR)$.

Deriving an approximation for $(\ES,\CovS)$ is also the goal of classic \emph{mean field} methods. In these methods, the magnetizations $\ES$ are approximated first, as the solution of some fixed point equation $\ES=F(\ES|\bh,\bJ)$. Then, this equation is differentiated wrt. $\bh$, yielding a predicted covariance matrix as
\begin{equation}
\label{eq:linresp}
C_{ij} = \partial_{h_i}m_j.
\end{equation}
Indeed, this so-called \emph{linear response} formula holds true in the exact Ising model~; so it provides a concrete way of estimating $\CovS$ from the approximation of $\ES$.

Informally, we may say that a given set of Ising parameters $(\bh,\bJ)$ lies in the `mean field domain' when \emph{some} mean field method can provide a good estimate of the corresponding moments $(\ES^\star,\CovS^\star)$. In the rest of this article, I will argue that the variational approximation of $P$ by a Cox distribution $Q$ is valid precisely in this `mean field domain'.

In this section, I briefly remind the nature of different mean field approximations, and compare their respective Plefka expansions in the case of weak couplings, up to order 3 (resp. 4 in the Appendices). In section \ref{sec:tests}, I will proceed to numerical comparisons.

\subsubsection*{Classic mean field approximations} 

I considered two classic mean field methods, the \emph{TAP} and \emph{Bethe} approximations, and a more recent generalization called the \emph{adaptive TAP} approximation. 
For the sake of self-completeness, these approximations are reminded in some detail in Appendices \ref{app:meanfield} and \ref{app:adaptiveTAP}. Here, I only provide essential formulas.

The archetypal mean field method in the Ising model is the \emph{TAP approximation}, where the approximating vector of magnetizations $\ES$ is sought as a fixed point of the following Thouless-Anderson-Palmer equation \cite{Thouless1977,Mezard1987,OpperSaad2001}~:
\begin{equation}
\label{eq:TAP}
m_{i} = \tanh\Big(h_i + \sum_{j\neq i} J_{ij}m_j - m_i \sum_{j\neq i} J_{ij}^2 (1-m_j^2)\Big).
\end{equation}
This equation can arise in different contexts, one of which is the Plefka expansion of the exact Ising model (eq. \ref{eq:ising3}), stopped at order 2.
Covariances $\CovS$ are derived in turn based on the linear response formula (\refeq{linresp})~:
\begin{align}
(\CovS^{-1}_{\rm TAP})_{ij} =& \left(1+\sum_k J_{ik}^2(1-m_i^2)(1-m_k^2)\right)\frac{\delta_{ij}}{1-m_i^2} \nonumber\\
& - J_{ij} - 2 J_{ij}^2m_im_j. \label{eq:TAP_linresp}
\end{align}

The \emph{Bethe approximation} is a related mean field method, where the approximating vector of magnetizations $\ES$ is sought as a fixed point of the following equation \cite{Yedidia2001,MezardParisi2001}~:
\begin{equation}
\tcavij = h_i + \sum_{k\neq j,i}\tanh^{-1}\big(\tanh(J_{ik})\tanh(\theta_k^{(i)})\big).
\label{eq:bethe}
\end{equation}
Here, the so-called \emph{cavity fields} $\tcavij$ are tractable functions of $\ES$, namely, the only numbers such that each 2-spin Ising distribution of natural parameters $(\tcavij,\tcavji,J_{ij})$ have first moments $(m_i,m_j)$. This equation can be justified as the exact solution for $\ES$ when the couplings $J_{ij}$ define a tree-like lattice, and its iterative resolution is known as the \emph{belief propagation} algorithm.
Covariances $\CovS$ can be derived in turn based on the linear response formula
(Appendix \ref{app:meanfield}, \refeq{bethe_linresp}, derived here with an original approach).

Finally, the \emph{adaptive TAP approximation} of Opper and Winther \cite{OpperWinther2001} is a generalized mean field approximation, based on the cavity method \cite{Mezard1987}. The magnetization variables $\{m_i\}$ are joined with a second set of variables $\{V_i\}$, which represent the variance of the cavity field distribution at each spin site, and obey the following fixed point equations~:
\begin{align}
m_i &= \tanh\big(h_i+\sum_j J_{ij}m_j - m_i V_i \big),\label{eq:opper_m} \\
(\CovS^{-1}_{\rm A})_{ij} &= \big(1+(1-m_i^2)V_i\big)\frac{\delta_{ij}}{1-m_i^2}- J_{ij}, \label{eq:opper_linresp}\\
1-m_i^2 &= (\CovS_{\rm A})_{ii}.\label{eq:opper_V}
\end{align}
This derivation is detailed in Appendix \ref{app:adaptiveTAP}. Briefly, \refeq{opper_m} is a generalization of the TAP equation \refno{TAP} where the variance $V_i$ of the cavity field is left as a free variable, and \refeq{opper_linresp} is the corresponding linear response prediction. Equation \refno{opper_V} imposes coherent predictions for individual variances $\Var(s_i)$, thereby closing the fixed point equation on variables $\{m_i,V_i\}$. 

The adaptive TAP approximation is a `universal' mean field method~: by letting the variances $V_i$ adapt freely, it can account for any statistical structure of matrix $\bJ$, whereas the classic TAP equation (\refeq{TAP}) is only true when the individual coupling weights $J_{ij}$ are decorrelated \cite{OpperWinther2001b}.
This is especially welcome in machine learning and neurosciences, where coupling strengths $J_{ij}$ are generally structured (because they represent learned regularities of the outside world).

\rev{%
Equations \refno{opper_m}-\refno{opper_V} bear a striking similarity with the variational Cox equations, \refeq{ERS}-\refno{CovS_var}. Both can be seen as modifications of the naive mean field equations through $N$ additional variables (the $d_i$, resp. $V_i$) associated to the \emph{variance} of the field acting on each spin. This similarity will be confirmed in the subsequent analytical and numerical results.
}

\subsubsection*{Weak coupling expansions}

In the Ising model, the link between natural parameters $(\bh,\bJ)$ and magnetizations $\ES$ can be abstractly described as $\bh=f(\ES,\bJ)$, for an intractable function $f$. Only when $\bJ=\mat{0}$ does the link become tractable~: the Ising model boils down to a Bernoulli distribution, with the obvious $h_i=\tanh^{-1}(m_i)$.

One step further, when couplings are weak but nonzero, one can derive the Taylor expansion of $f$ around $\bJ=\mat{0}$. In practice, the coupling matrix is written $\alpha\bJ$, $\alpha$ being the small parameter of the expansion, and the result is known as the \emph{Plefka expansion} \cite{Plefka1982} -- although the approach can be traced back to anterior work \cite{VasilevRadzhabov1974}.

The expansion up to order 4 is a classic computation, outlined in Appendix \ref{app:meanfield}. Stopping at order 3 for brevity, it reads~:
\begin{widetext}
\begin{align}
\label{eq:ising3}
h_i &= \tanh^{-1}(m_i) - \alpha\sum_{j\neq i}J_{ij}m_j + \alpha^2 m_i\sum_{j\neq i}J_{ij}^2(1-m_j^2) \nonumber\\&
+\alpha^3\left[2m_i\sum_{(jk|i)}J_{ij}J_{jk}J_{ki}(1-m_j^2)(1-m_k^2) + 2(m_i^2-\tfrac{1}{3})\sum_{j\neq i} J_{ij}^3m_j(1-m_j^2)\right]
+o(\alpha^3),
\end{align}
where $(jk|i)$ denotes all unordered triplets 
of the form $\{i,j,k\}$ with $j$ and $k$ distinct, and distinct from $i$.
\end{widetext}

This expansion can serve as a first test on the various approximations (TAP, Bethe, adaptive TAP, variational Cox) introduced above. Indeed, these approximations can also be described as $\bh=f(\ES,\bJ)$ for a different function $f$, and we can compare its Taylor expansion to that of the true Ising model.

The expansion for the TAP approximation is, by definition, the exact Ising expansion (\refeq{ising3}) stopped at order 2. The expansion for the Bethe approximation is obtained from the exact Ising expansion by retaining only the sums over spin pairs \cite{Ricci2012} (see Appendix \ref{app:meanfield}). In \refeq{ising3}, this means suppressing the sum over $(jk|i)$ in the order 3 term, but keeping the sum over $j\neq i$.

The expansion for the adaptive TAP approximation, derived in Appendix \ref{app:adaptiveTAP}, writes
\begin{widetext}
\begin{align}
\label{eq:adap3}
h_i &= \tanh^{-1}(m_i) - \alpha\sum_{j\neq i}J_{ij}m_j + \alpha^2 m_i\sum_{j\neq i}J_{ij}^2(1-m_j^2) \nonumber\\&
+\alpha^3\Big[2m_i\sum_{(jk|i)}J_{ij}J_{jk}J_{ki}(1-m_j^2)(1-m_k^2)\Big]
+o(\alpha^3).
\end{align}
The expansion for the variational Cox approximation, derived in Appendix \ref{app:perturb}, writes
\begin{align}
\label{eq:cox3}
h_i &= \tanh^{-1}(m_i) - \alpha\sum_{j\neq i}J_{ij}m_j + \alpha^2 m_i\sum_{j\neq i}J_{ij}^2(1-m_j^2) \nonumber\\&
+\alpha^3\Big[2m_i\sum_{(jk|i)}J_{ij}J_{jk}J_{ki}(1-m_j^2)(1-m_k^2) - 2(m_i^2-\tfrac{1}{3})J_{ii}^3m_i(1-m_i^2)\Big]
+o(\alpha^3).
\end{align}
\end{widetext}

At order 2, all approximations considered have the same expansion as the exact Ising solution, meaning that they will perform well in case of weak couplings.

At order 3, discrepancies appear between the exact Ising solution and its various approximations. The first contribution to the order 3 term in \refeq{ising3}, the sum over $(jk|i)$, is correctly accounted for by the adaptive TAP and variational Cox approximations. The second contribution to the order 3 term in \refeq{ising3}, the sum over $j\neq i$, is correctly accounted for by the Bethe approximation. Of these two sums, that over $(jk|i)$ involves many more terms, so we expect it will generally be the dominant contribution, except for very specific coupling matrices $\bJ$. 

At order 4, the same qualitative features are observed. The `generally dominant' contribution at order 4 in the true Ising solution writes
\[
2m_i \sum_{(jkl|i)}J_{ij}J_{jk}J_{kl}J_{li}(1-m_j^2)(1-m_k^2)(1-m_l^2)
\]
(Appendices, \refeq{ising4}), and this is also the dominant contribution to the adaptive TAP (\refeq{adap4}) and variational Cox (\refeq{cox4}) approximations. Hence, we may expect these two approximations to provide a better fit than the others in case of generic coupling matrices $\bJ$ -- and this will indeed be our observation in numerical tests (Figure \ref{figk}).

The similar structures of \refeq{CovS_var} and \refno{opper_linresp} suggest a proximity between the adaptive TAP and variational Cox approximations, and this is confirmed by their weak coupling expansions~: up to order 4, their respective expansions differ only through additional terms involving the diagonal weights $J_{ii}$, so they would be identical for a classic coupling matrix such that $\diag(\bJ)=\mat{0}$.

\section{Numerical tests}
\label{sec:tests}

To gain more insights on the behavior of all approximations above, I turned to numerical exploration~: I picked a large number of possible configurations $(\bh,\bJ)$, estimated the true Ising moments $(\ES^\star,\CovS^\star)$ in each configuration with lengthy MCMC sampling, and compared all approximations against this ground truth.
The numerical details for computing the approximations are given in Appendix \ref{app:numerical}.

\rev{I should stress from the start that the role of these tests is \emph{not} to target the most accurate mean field method for approximating $(\ES,\CovS)$ -- which turns out to be the adaptive TAP method, in most configurations tested here. Instead, the goal of these tests is to support one main claim of this article~: that the Ising distribution is well approximated by a Cox distribution in the same domain of parameters where mean field methods are efficient.}

To focus on the most important aspect of the Ising model, I only tested configurations where $\bh=\mat{0}$, so that magnetizations verify $\ES=\mat{0}$ -- both in the exact Ising solution and in the various approximations. Hence, the efficiency of a given approximation is assessed by its ability to correctly predict the covariance matrix $\CovS$.
For each tested configuration and approximation, the fit performance is summarized by number
\begin{equation*}
z = \frac{1}{N^2}\sum_{i,j} \big|C_{ij} - C_{ij}^\star\big|,
\end{equation*}
where $C_{ij}^\star$ is the true covariance of spins $i$ and $j$, and $C_{ij}$ its approximation.

\subsubsection*{Generative model for couplings $\bJ$}

In this approach, the choice of a generative model for coupling matrix $\bJ$
 is a delicate matter, as the Ising model can exhibit very different behaviors depending on its parameters. 
 To test different regimes with a single formula, I used the following generative model~:
\begin{align}
\bJ &=\left[\frac{J_0}{pN}\mathbf{1} + \frac{J}{\sqrt{\kappa p} N}\mat{X}_\kappa\mat{X}_\kappa^\top\right] \;  .^* \; \mat{M}_p  \label{eq:SKgen}
\end{align}
with $\mathbf{1}$ the $N\times N$ matrix with uniform unit entries and $\mat{X}_\kappa$ an $N\times \kappa N$ matrix of independent standard normal entries. Afterwards, element-wise multiplication by a symmetric masking matrix $\mat{M}_p$ can randomly set each edge $(ij)$ at 0 with probability $1-p$.

This model has 4 parameters. The positive numbers $(J_0,J)$ correspond to the standard Sherrington-Kirkpatrick (SK) parameters for spin glasses \cite{SherringtonKirkpatrick1975}, meaning that the probabilistic distribution of each nonzero off-diagonal term $J_{ij}$ follows
\begin{equation}
J_{ij}\sim \mathcal{N}\left(\frac{J_0}{pN}, \frac{J^2}{pN}\right)
\label{eq:SK}
\end{equation}
to a very good approximation, owing to the central limit theorem applied to the $\kappa N$ samples in matrix $\mat{X}_\kappa$.

Parameter $\kappa>0$ fixes the amount of global correlation between individual couplings $J_{ij}$.
When $\kappa\rightarrow+\infty$, all off-diagonal entries $J_{ij}$ constitute independent random variables, and the classic SK model is recovered (see Supplementary Material).
When $\kappa$ is smaller, the random matrix $\mat{X}_\kappa\mat{X}_\kappa^\top$ follows a Wishart distribution, as in the Hopfield model of associative memory \cite{hopfield1982,Mezard1987}. The random variables $J_{ij}$ become dependent, and the spectrum of $\bJ$ differs markedly from the SK case (Wigner vs. Mar\v{c}enko-Pastur laws). Note that the probabilistic distribution of each element $J_{ij}$ remains virtually unchanged in the process, given by \refeq{SK} except at very low values of $\kappa$.

Finally, parameter $p\in[0,1]$ allows to dilute the overall connectivity, so that only a proportion $p$ of the couplings are nonzero. Coherently, $J_0$ and $J^2$ in \refeq{SK} are scaled by $pN$, the effective number of neighbors in the (possibly diluted) model.

\begin{figure}
\includegraphics[width=\linewidth]{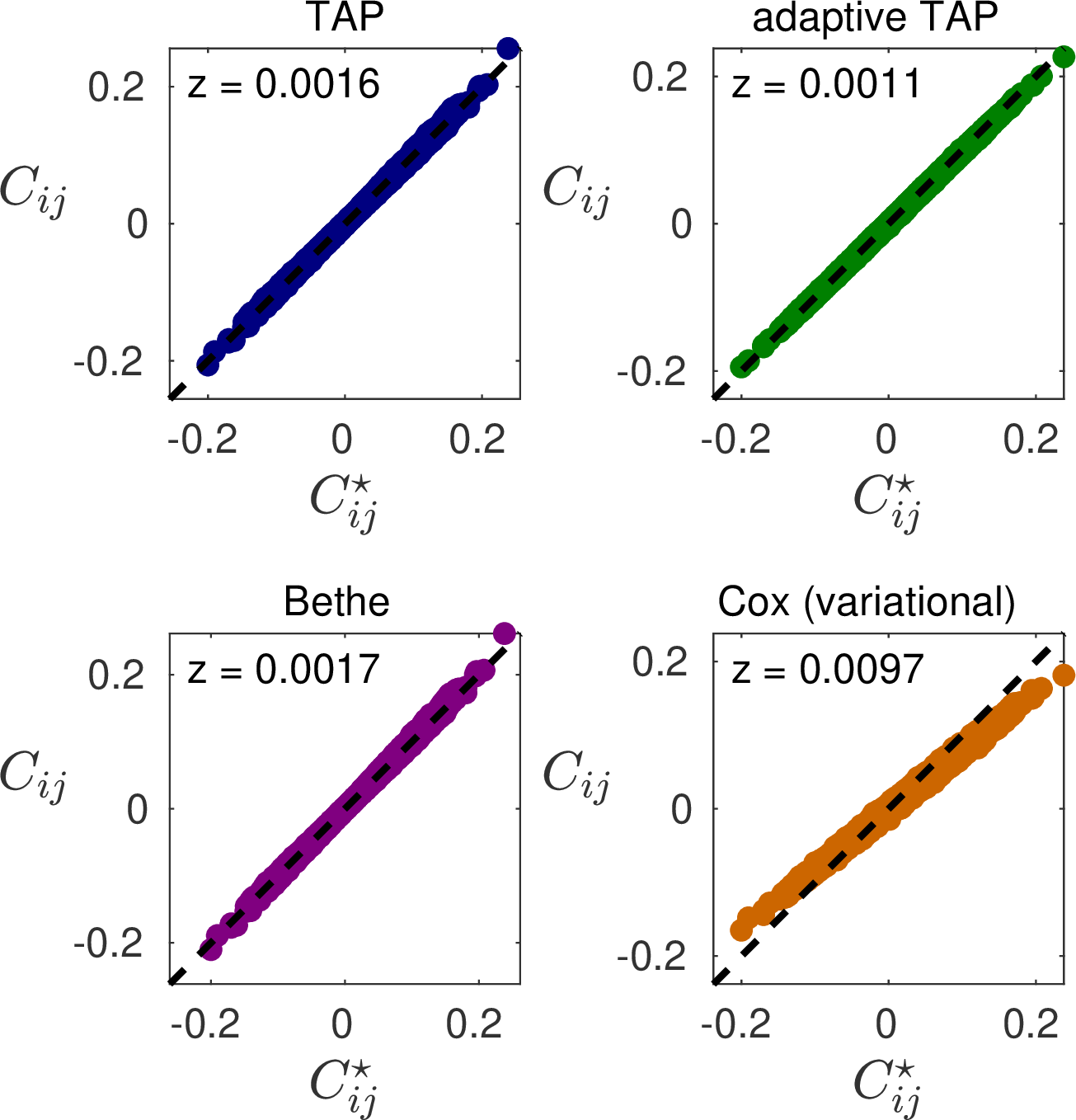}
\caption{\label{figbase}
TAP, Bethe, adaptive TAP and variational Cox approximations on a typical paramagnetic configuration with $N=100$ spins, and reference parameter values $(J_0,J,\kappa,p)=(0.5,0.5,\infty,1)$. The corresponding fit values $z$ are indicated as insets.
In each of Figures \ref{figJ0}-\ref{figk}, one parameter will be modified, while the three others keep their reference value.
}
\end{figure}

\subsubsection*{Results for the approximations}

As the 4 parameters in model \refno{SKgen} prevent from an exhaustive search, I performed a restricted exploration of parameter space, based on a set of reference parameter values~:
\begin{equation}
\label{eq:refpar}
(J,J_0,\kappa,p) = (0.5,0.5,\infty,1).
\end{equation}
Since $(\kappa,p)=(\infty,1)$, the individual coupling weights $J_{ij}$ are drawn independently and the connectivity matrix is dense. Hence, this is a classic SK model in its paramagnetic phase, because $J$ and $J_0$ are smaller than 1 \cite{SherringtonKirkpatrick1975}.

Figure \ref{figbase} shows the fit performances of the various approximation methods on a typical configuration $(\bh,\bJ)$ with these generative parameters. All approximations perform well, as expected, since the TAP equations are exact when $N\rightarrow\infty$ in the paramagnetic phase of the SK model \cite{Mezard1987}.
However, the variational Cox approximation (fourth panel) displays a bias~: the overall magnitude of its predictions $C_{ij}$ is somewhat underestimated, leading to a slant in the graph of $(C_{ij}^\star,C_{ij})$, and a larger fit value $z$.
This bias is a systematic property of the Cox approximation, which is mainly caused by the presence of nonzero self-coupling terms $J_{ii}>0$ (see Discussion).

\begin{figure}
\includegraphics[width=\linewidth]{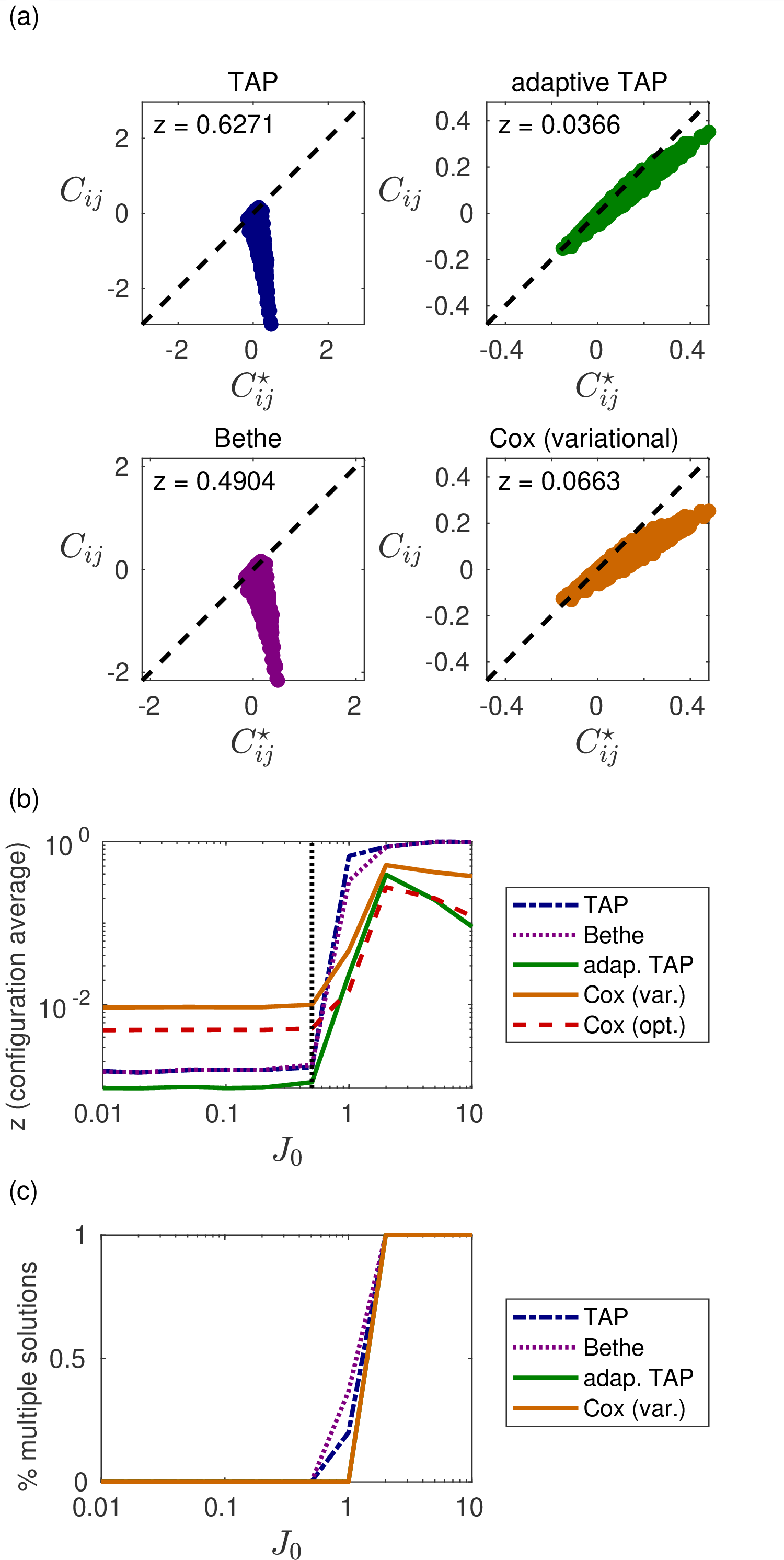}
\caption{\label{figJ0}
Transition from paramagnetic to ferromagnetic phase, when parameter $J_0$ is increased. Other parameters keep their reference value $(J,\kappa,p)=(0.5,\infty,1)$.
(a) TAP, Bethe, adaptive TAP and variational Cox approximations on a typical configuration with $J_0=1$.
(b) Average fit value $z$ for each of the approximations, assessed over 30 configurations of matrix $\bJ$ picked according to \refeq{SKgen}, for each tested value of $J_0$. Fit measure is also provided for the optimal Cox distribution (see text). The vertical dashed line corresponds to the reference values of Figure \ref{figbase}.
(c) Apparition of multiple solutions to the respective fixed point equations. For each value of $J_0$ and approximation considered, I plot the percentage of the 30 configurations in which multiple solutions were found (see Appendix \ref{app:numerical}, \emph{Numerical procedures}).
}
\end{figure}

\begin{figure}
\includegraphics[width=\linewidth]{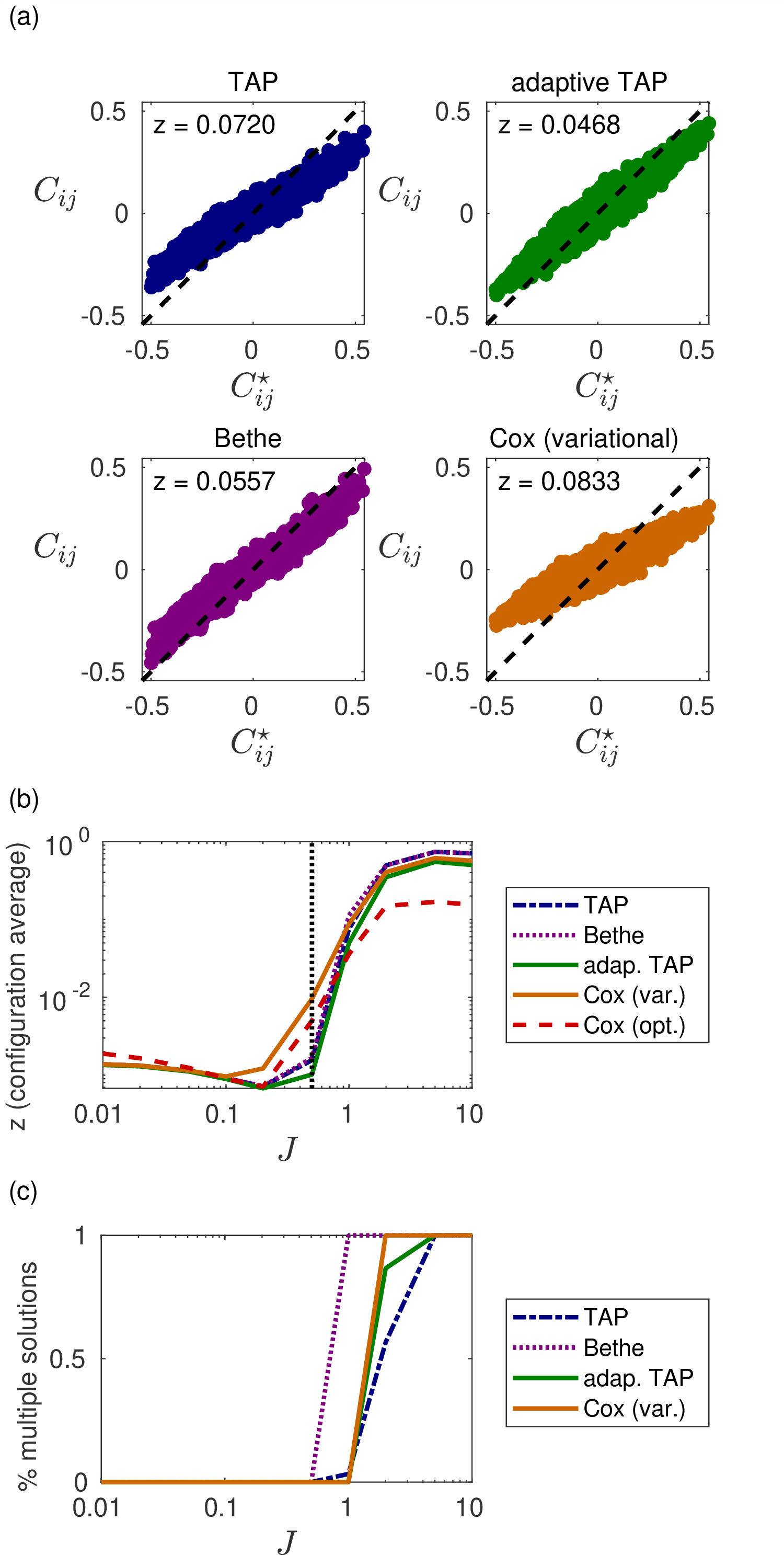}
\caption{\label{figJ}
Transition from paramagnetic to spin glass phase, when parameter $J$ is increased. Other parameters keep their reference value $(J_0,\kappa,p)=(0.5,\infty,1)$. 
(a) TAP, Bethe, adaptive TAP and variational Cox approximations on a typical configuration with $J=1$.
(b), (c)~: same as Figure \ref{figJ0}.
}
\end{figure}

\begin{figure}[t]
\includegraphics[width=\linewidth]{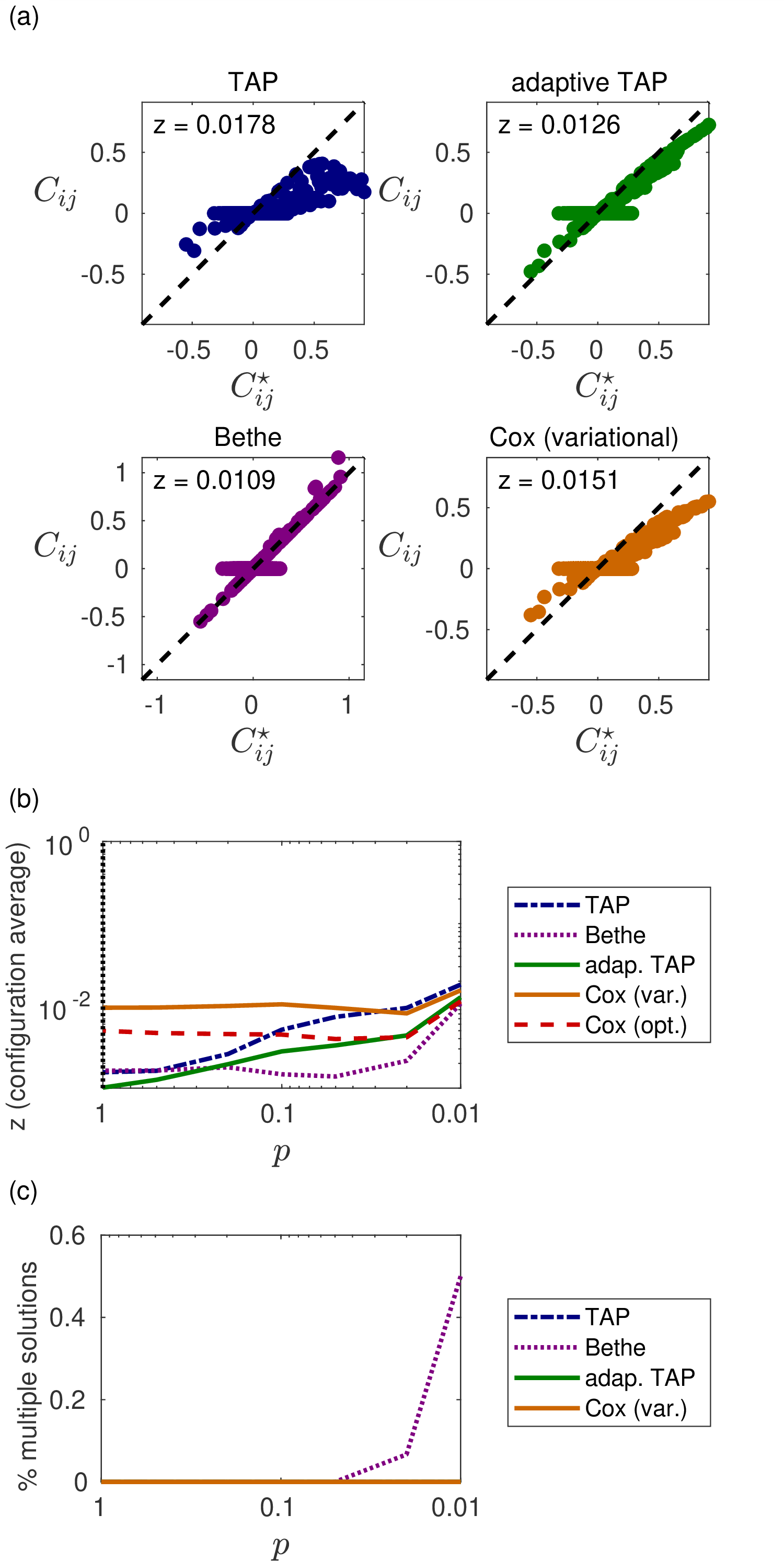}
\caption{\label{figp}
Introduction of sparseness in the coupling weights, when parameter $p$ is decreased. Other parameters keep their reference value $(J,J_0,\kappa)=(0.5,0.5,\infty)$. 
(a) TAP, Bethe, adaptive TAP and variational Cox approximations on a typical configuration with $p=0.01$.
(b), (c)~: same as Figure \ref{figJ0}.
}
\end{figure}

\begin{figure}
\includegraphics[width=\linewidth]{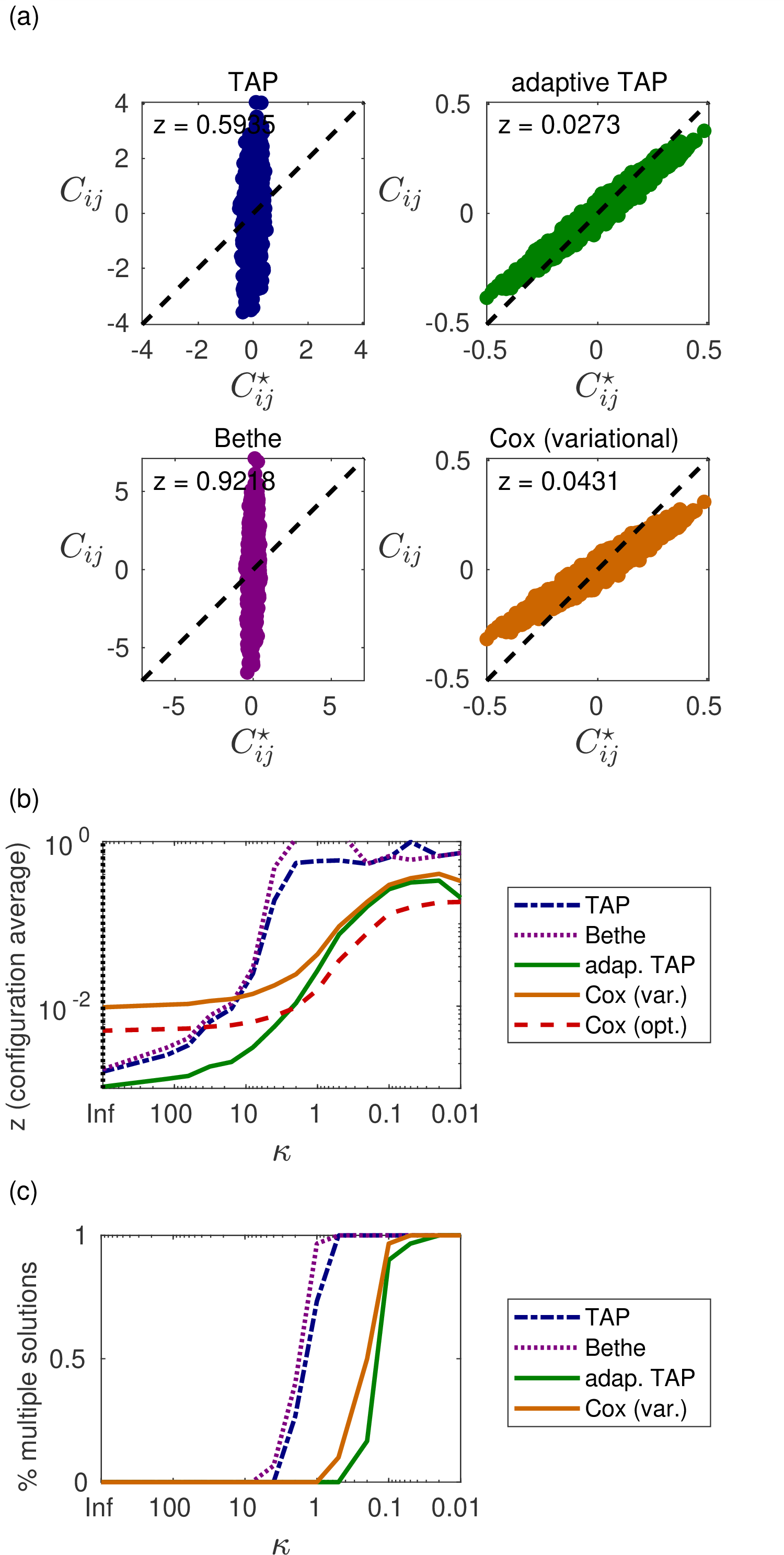}
\caption{\label{figk}
Introduction of stochastic dependence between the coupling weights, when parameter $\kappa$ is decreased. Other parameters keep their reference value $(J,J_0,p)=(0.5,0.5,1)$. 
(a) TAP, Bethe, adaptive TAP and variational Cox approximations on a typical configuration with $\kappa=1$.
(b), (c)~: same as Figure \ref{figJ0}.
}
\end{figure}

I then explored the approximations' behavior in different departures from the SK paramagnetic situation.
In each of Figures \ref{figJ0}-\ref{figk}, one parameter in \refeq{refpar} is varied while the three others keep their reference value. 
Panel (a) shows the various approximations on a typical configuration at the transition out of the `paramagnetic SK' phase.
Panel (b) shows the mean fit performance of each approximation as the concerned parameter is varied.
Panel (c) shows when multiple solutions have been detected to each approximation's constitutive fixed point equation.


I first tested the approximations' behavior when transiting into the ferromagnetic (Figure \ref{figJ0}) and spin-glass (Figure \ref{figJ}) phases of the SK model. The ferromagnetic phase, corresponding to $J_0>{\rm max}(1,J)$, is characterized by a symmetry breaking into two `magnetized' states with $s_i=1$ (resp. $-1$) for all spins, constituting the stable solutions of the TAP equation \cite{Mezard1987,Nishimori2001}.
The spin glass phase, corresponding to $J>{\rm max}(1,J_0)$, is characterized by the apparition of multiple `metastable' local minima of the TAP free energy, with limited basins of attraction \cite{Mezard1987}.
All approximations considered have roughly the same behavior at the phase transitions (Figure \ref{figJ0}, $J_0\geq 1$, Figure \ref{figJ}, $J\geq 1$), losing precise fit (panels (b)) concurrently with the apparition of multiple solutions to their respective equations (panels (c)). This confirms the existence of a universal `mean field' domain for the SK model, corresponding to its paramagnetic phase.
\footnote{%
Looking in more detail, a notable qualitative difference exists between the different approximations, at least in the ferromagnetic phase.
In the TAP and Bethe approximations, the symmetric solution with $\ES=\mat{0}$ becomes unstable \cite{Mezard1987,Nishimori2001}, and the linear response prediction for $\CovS$ at this point diverges -- as visible in Figure \ref{figJ0}(a). In the adaptive TAP and variational Cox approximations, this symmetric solution remains stable and coexists with the magnetized solutions (as in Figure \ref{fig2D}(b)), and the prediction for $\CovS$ is progressively degraded, rather than totally lost -- see Figure \ref{figJ0}(a).}

When couplings are made sparser, all approximations again display the same qualitative behavior (Figure \ref{figp}), maintaining a reasonable precision down to very diluted models. The Bethe approximation is the most efficient in this case, because the rarefaction of loops creates a `tree-like' structure of connectivity.

The main difference between the approximations is their handling of structured coupling matrices $\bJ$ (Figure \ref{figk}). When parameter $\kappa$ decreases and couplings weights $J_{ij}$ become correlated, the TAP and Bethe approximations deteriorate much faster than the variational Cox and adaptive TAP approximation -- which was designed precisely for this purpose \cite{OpperWinther2001}.

This numerical study confirms the adaptive TAP's interest as a `universal' mean-field method, the most efficient in all tested regimes with dense couplings, and second most efficient in case of sparse couplings (the Bethe approximation performing marginally better). It also reveals similar domains of validity for the adaptive TAP and variational Cox approximations -- notwithstanding the latter's systematic bias in `easy' configurations, leading to higher fit values $z$.
In summary, a `mean field domain' can be defined as the ensemble of parameters $(\bh,\bJ)$ for which the adaptive TAP method efficiently predicts the spin moments $(\ES,\CovS)$, and this is also the domain where the Ising distribution can be easily replaced by a Cox approximation, thanks to a variational principle.

\subsubsection*{Optimal Cox distribution}

\rev{%
As such, the above results do not explicitly tell whether the Ising distribution can be approximated by a Cox distribution outside of the `mean field domain'. It may be the case that a decent Cox approximation exists, but cannot be retrieved by a variational principle anymore.
To clarify this point, I also considered the fit performance of the \emph{optimal} Cox distribution $Q(\ER^\star,\CovR^\star)$ defined above. My measure of fit in this case consisted in comparing the true spin covariances $C_{ij}^\star$ to their values in the optimal Cox distribution, as given by \refeq{CovScox}. 
Thus, a successful fit indicates when assuming a Gaussian shape for the latent field distribution $P(\br)$, without modifying its moments, does not modify much the resulting spin moments.

Figures \ref{figJ0}-\ref{figk}(b) show that this measure globally correlates with the efficiency of the adaptive TAP and variational Cox approximations, i.e, it also deteriorates outside of the `mean field domain'. Hence, the increased discrepancy between Ising and Cox distributions outside of the `mean field domain' 
seems intrinsically related to the Ising latent field $P(\br)$ becoming non-Gaussian.
This is also coherent with the fact that the adaptive TAP approximation is bound to fail precisely when the instantaneous field acting on each spin cannot be considered Gaussian (see Appendix \ref{app:adaptiveTAP}).

\begin{figure*}
\includegraphics[width=\linewidth]{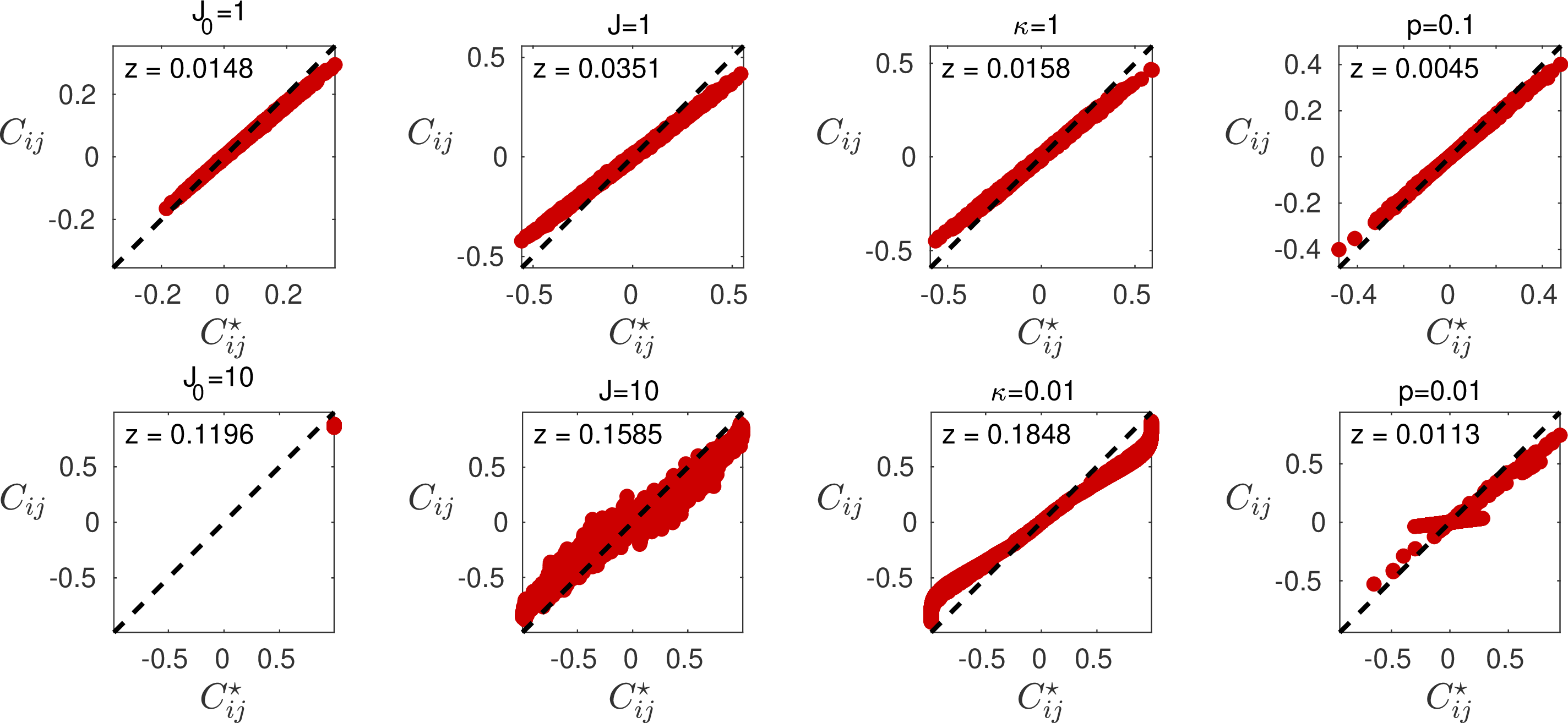}
\caption{\label{figopt}
Fit for the optimal Cox distribution on typical configurations at the boundary (first row) and far outside (second row) of the paramagnetic SK phase. 
Each panel plots the pairwise spin covariances in the true Ising distribution ($C_{ij}^\star$) against their values in the optimal Cox distribution ($C_{ij}$), computed with \refeq{CovScox}. 
Starting from reference parameter values $(J_0,J,\kappa,p)=(0.5,0.5,\infty,1)$, the approximation is assessed at different values of $J_0$ (first column, transition to ferromagnetic SK model), $J$ (second column, transition to spin glass SK model), $\kappa$ (third column, introduction of correlations between coupling weights $J_{ij}$), and $p$ (fourth column, dilution of connectivity).
}
\end{figure*}

Nonetheless, in quantitative terms, the loss of fit by the optimal Cox distribution is never total.
Figure \ref{figopt} shows examples of fit performance for the optimal Cox distribution in various configurations at the boundary (first row, compare to Figures \ref{figJ0}-\ref{figk}(a)) and far outside (second row) of the mean field domain. It reveals that, even when the Ising latent field $P(\br)$ is far from being Gaussian-distributed, its replacement by a Gaussian preserves the overall pattern of spin correlations (as visible in Figure \ref{fig2D}(b)). This global correctness is hardly reflected in the magnitude of error $z$, yet it does imply that an Ising model is never `too far' away from its optimal Cox distribution.

}

\section{Discussion}

I have proposed a reformulation of the Ising distribution as a latent variable model, and used it to derive principled approximations by the simpler Cox distribution.
In practical applications, Cox models (including the dichotomized Gaussian, and Cox point processes) are often preferred to the corresponding maximum entropy distributions (Ising distribution, Gibbs point process) because they are easier to sample, and to parametrize from a set of observed moments. This article establishes a simple analytical connection between the two families of models, and investigates under what conditions they can be used interchangeably.

The most natural connection between an Ising and a Cox distribution 
is obtained by equating the two first moments of their latent variables, \refeq{ERSstar}-\refno{CovRSstar}, a simple but fundamental result of the article. The resulting `optimal' Cox approximation holds well in paramagnetic conditions, and even beyond, as far as global trends are concerned (Figure \ref{figopt}). However, \refeq{ERSstar}-\refno{CovRSstar} involve both the natural parameters $(\bh,\bJ)$ and the resulting moments $(\ES^\star,\CovS^\star)$ of the Ising distribution, which makes them unpractical in most concrete situations.

\rev{%
To target a Cox approximation given only some natural parameters $(\bh,\bJ)$, I have explored a classic variational approach, leading to \refeq{ERS}-\refno{CovS_var}.
These equations are not particularly interesting as a mean-field method for predicting $(\ES,\CovS)$, since they globally behave as a biased version of Opper and Winther's adaptive TAP method. However, their analytical (Section \ref{sec:meanfield}) and numerical (Section \ref{sec:tests}) analysis allows to formulate the key conclusion of this article~: mean-field methods are efficient precisely when the Ising distribution can be associated to a quasi-normal latent field distribution.

If the practical goal is to establish a Cox approximation for the Ising distribution of parameters $(\bh,\bJ)$, the moments $(\ES,\CovS)$ may as well be estimated with any other choice of mean field method, and then input to \refeq{ERS}-\refno{CovRS} to produce the corresponding Cox approximation.}

Naturally, the relation to mean field methods is not accidental. From \refeq{RcondS}, given a spin configuration $\bs$, the latent field $r_i$ is distributed as $\Norm(h_i+\sum_j J_{ij}s_j,J_{ii})$. In particular, if self-couplings $J_{ii}$ are zero as in the classic Ising model, we simply recover
\[
r_i = h_i + \sum_{j\neq i} J_{ij}s_j,
\]
that is, the instantaneous field variable considered in the cavity method (Appendix \ref{app:adaptiveTAP}, \refeq{hcav}) and adaptive TAP equations.
Thus, the latent field formalism differs from the classic cavity approach only through the role of nonzero self-couplings $J_{ii}$. Coherently, the weak coupling expansions of the variational Cox and adaptive TAP approximations up to order 4 differ only because of nonzero $J_{ii}$. 

This suggests that nonzero self-couplings $J_{ii}$ may be responsible for the systematic bias of the Cox approximation in its prediction of $(\ES,\CovS)$, compared to the adaptive TAP approximation. 
As a direct confirmation, I observed that when matrix $\bJ$ is given a zero diagonal, the variational Cox equation \refno{ERS}-\refno{CovS_var} generally retains a solution, and it is then remarkably close to the adaptive TAP solution (see Supplementary Material). 
Unfortunately, nonzero weights $J_{ii}$ are required to endow the field variables $r_i$ with a true, multivariate distribution $P(\br)$ (this is not the case in the cavity method), and thus produce a \emph{concrete} approximation of the Ising distribution by a simpler latent variable distribution.

\rev{%
A disturbing consequence is that there is not one latent field distribution associated to the Ising distribution, but \emph{many} different distributions, depending on the value given to self-couplings $J_{ii}$. When all values $J_{ii}$ are taken very large, the latent field $P(\br)$ is `useless'~: it is the mere mixture of $2^N$ Gaussian bumps with no overlap, located on the $2^N$ summits of a hypercube, and the bump at summit $\bs$ is simply associated to weight $P(\bs)$  (as in Figure \ref{fig2D}(b)).
Then, as the $J_{ii}$ become smaller, some of the $2^N$ bumps start overlapping, and $P(\br)$ acquires a less trivial overall distribution. In some cases, when the $J_{ii}$ are made small enough, the overall shape of $P(\br)$ becomes quasi-Gaussian (as in Figure \ref{fig2D}(a)). In other cases, $P(\br)$ never becomes quasi-Gaussian, because a lower bound is reached where the $J_{ii}$ cannot be made smaller while ensuring that matrix $\bJ$ remains definite positive.

Given some classic Ising parameters $\{J_{ij}\}_{i<j}$, the set of self-couplings $J_{ii}$ obtained with \refeq{TrJ} represents the ``smallest'' diagonal elements that can be used, and we could suppose that if the latent field $P(\br)$ is still not Gaussian for these values, it won't be either for other values of $J_{ii}$. Interestingly, the optimization problem in \refeq{TrJ} has an intrinsic significance for the `classic' Ising distribution (without self-couplings)~: the resulting value of $-\sum_i J_{ii}$ constitutes the Lagrangian dual approximation for the minimum of $\sum_{i<j} J_{ij}s_is_j$ over all $2^N$ spin configurations \cite{Boyd2004}, that is, the log-likelihood of the most unlikely spin configuration in the model. Arguments of this type suggests that, while self-couplings $J_{ii}$ are extraneous elements to the standard Ising model, the \emph{set of admissible values} for the $J_{ii}$ may have theoretical links with the nature and global difficulty of the considered Ising model.

Empirically, I observed a certain robustness to the exact choice of $\diag(\bJ)$, and the values could generally be doubled without affecting much the numerical results. In some cases, increasing all self-couplings $J_{ii}$ a little from the `optimal' solution of \refeq{TrJ} can even improve the fit performance of the variational Cox approximation, presumably because it reduces the condition number of matrix $\bJ$ (see also Appendix \ref{app:numerical}).
Thus, theoretically as well as practically, the significance and optimal choice of $\diag(\bJ)$ is not totally settled, and could be the subject of future work.
}

More generally, the present work could give rise to a number of developments.
For example, the theoretical study of phase transitions in the variational Cox approximation, as observed in Figures \ref{figJ0}-\ref{figk}(c), remains to be done.
Intuitively, these phase transitions are related to the apparition of multiple, well-separated modes in the distribution of the latent fields (see Figure \ref{fig2D}), which would then play a role similar to the ``pure states'' of spin glass theory \cite{Mezard1987,Nishimori2001}.
 
On a more applied level, the latent field variables $r_i$ could be incorporated into MCMC sampling schemes for the Ising distribution. For example, drawing the initial spins $s_i$ with a Cox approximation $Q(\bs)$, instead of a classic independent Bernoulli draw, can largely reduce the chain's convergence time to its equilibrium distribution $P(\bs)$. One step further, one could devise MCMC schemes that directly sample the fields' distribution $P(\br)$ and use it to generate the spins.

Finally, the formalism of latent variables could be applied to the inverse problem of retrieving $(\bh,\bJ)$ from a set of observed moments $(\ES,\CovS)$, which is arguably the biggest obstacle in practical applications of the Ising model listed above \cite{Weigt2009,Schneidman2006,Ohiorhenuan2010}.
On the one hand, advanced methods of mean-field inspiration have been developed in the last decade to tackle the inverse Ising problem \cite{Sessak2009,Roudi2009,Decelle2016}. 
On the other hand, it has been suggested to replace the Ising distribution by a dichotomized Gaussian (a limiting case of Cox distribution) in practical applications, precisely because it offers an easier inverse problem \cite{Macke2011}. 
Hopefully, the latent field formalism can reconcile the two approaches in a unified picture.


\newpage

\appendix

\section{Numerical procedures}
\label{app:numerical}

In this appendix, I give numerical details of the tests presented in Section \ref{sec:tests}.
All approximation methods considered are naturally described by a fixed point equation of the form $X=F(X)$. A simple heuristic for solving such equations is a numerical scheme 
\[
X_{n+1}=(1-\alpha)X_n+\alpha F(X_{n}),
\]
with $\alpha$ a small, adaptive, update parameter. When $\|X_{n}-F(X_n)\|$ is found to increase between two successive iterations, $\alpha$ is divided by 10. Else, $\alpha$ is multiplied by $1.05$.
 This heuristic proved sufficient to target a fixed point, in all cases encountered.
 
In the Cox approximation, the variable was $X=\{m_i,d_i^{-1}\}_{i=1\dots N}$. In the adaptive TAP approximation, it was $X=\{m_i,V_i\}_{i=1\dots N}$. The starting point $X_0$ was chosen with $m_i=0$, and numbers $\Delta_i=d_i^{-1}$ (resp $\Delta_i=1+V_i$) as the minimal values ensuring that matrix $\CovS^{-1}=\diag(\Delta_i)-\bJ$ be definite positive (see \refeq{CovS_var} and \refeq{opper_linresp}).

In the TAP and Bethe approximations, the starting point $X_0=\ES=\mat{0}$ was directly a solution of the fixed point equation, and I simply used the linear response prediction for $\CovS$ at this point. This seemed the fairer choice in the context of these tests, even though in some parameter regimes (e.g., SK ferromagnetic phase) the solution at $\ES=\mat{0}$ is unstable~: the iterative scheme started at any neighboring point does not converge to this solution.
 
It is well-known that, in certain regimes of parameters, the Ising model can display spontaneous symmetry breaking \cite{Mezard1987}. First, ergodicity breaking can occur in the MCMC chain used to sample the distribution. To counteract this effect, the true moments $(m_i^\star,C_{ij}^\star)$ were estimated from several independent MCMC chains with simulated annealing. 

Second, the various approximations themselves can start displaying multiple solutions to their constitutive equation \cite{Mezard1987}. To assess this effect, for each tested configuration $(\bh,\bJ)$ and approximation formula, I relaunched the numerical search from different starting points, namely, the empirical means 
found in each individual MCMC chain used during the sampling phase. This procedure only served as a (rudimentary) attempt to detect the presence of multiple solutions. It did not affect the measure of fit for the approximation, which was always based on the solution found from the starting points $X_0$ listed above.

In the Cox approximations (optimal and variational), for each configuration with off-diagonal elements $\{J_{ij}\}_{i<j}$, the diagonal couplings $J_{ii}$ were chosen as the solution of \refeq{TrJ}, plus a constant ridge term $\lambda$ chosen for $\bJ$ to have a condition number of $10$. Indeed, ill-conditioned matrices $\bJ$ lead to increased errors in some parameter regimes.

\bigskip

\section{Variational mean field approximations}
\label{app:meanfield}


In this appendix, I recall the variational approach of mean field theory, which can be used to recover the Plefka expansion of the exact Ising model, as well as the TAP and Bethe approximations.

\subsubsection*{Mean field variational approach}

As in the main text, let us note $P(\bs)$ the true Ising distribution of parameters $(\bh,\bJ)$, and $(\ES^\star,\CovS^\star)$ its corresponding moments. Let $Q(\bs)$ any other distribution proposed as an approximation of $P$. The KL divergence ${\rm KL}(Q||P)=\sum_{\bs}Q(\bs)\ln(Q(\bs)/P(\bs))$ can be expressed as
\begin{align*}
{\rm KL}(Q||P)
&= U(Q) - S(Q) +\ln Z_I(\bh,\bJ),
\end{align*}
where $U(Q)$ is the average Ising energy, and $S(Q)$ the entropy, under distribution $Q$~:
\begin{align*}
U(Q) &:= -\sum_{\bs}Q(\bs)\Big(\bh^\top\bs+\frac{1}{2}\bs^\top\bJ\bs\Big),\\
S(Q) &:= -\sum_{\bs}Q(\bs)\ln Q(\bs).
\end{align*}
The functional $G(Q):=U(Q)-S(Q)$ is called the \emph{variational (or Gibbs) free energy} of distribution $Q$ as an approximation of $P$. Smaller values of $G(Q)$ correspond to a lower KL divergence and thus to a better fit, and its minimum is achieved for $Q=P$.


In the Ising distribution, given fixed couplings $\bJ$, there is a one-to-one correspondence between values of $\bh$ and resulting values of $\ES=\pE(\bs)$.
Thus, in theory, we can apply the variational approach to the family of Ising distributions $Q(\ES,\bJ)$, where couplings $\bJ$ are taken equal to those in $P$, and $\ES$ constitutes the $N$-dimensional parametrization variable. The natural field parameters of $Q$, say $\btheta$, are then a (generally intractable) function of $(\ES,\bJ)$, and the associated free energy writes
\begin{equation}
\label{eq:Gmf}
G(\ES,\bJ)= (\btheta-\bh)^\top \ES - \ln Z_I(\btheta,\bJ),
\end{equation}
whose minimum over $\ES$ is obtained when $\ES=\ES^\star$, that is, when $Q=P$. (Note that $G$ also depends on the field parameter $\bh$ of distribution $P$, but I omit it for lighter notations.)

Using the classic conjugacy relation $\partial_{\btheta}\ln Z_I = \ES$ in the Ising model, one can note that
\begin{equation}
\label{eq:Gmf_m}
\partial_{\ES} G(\ES,\bJ) = \btheta-\bh,
\end{equation}
so function $G(\ES,\bJ)+\bh^\top\ES$ corresponds to the Legendre transform of $\ln Z_I$ in its first variable.

Function $G(\ES,\bJ)$ is not tractable in general, but it can be approximated -- and the resulting minimum will yield an approximation of the true moments $\ES^\star$. This approach is known as the \emph{mean field} variational method, of which the TAP and Bethe approximations are two prominent examples.

\subsubsection*{TAP and Plefka approximations}

In so-called \emph{Plefka expansions}, one approximates $G(\ES,\bJ)$ by its Taylor expansion in $\bJ$ around $\bJ=\mat{0}$~:
\begin{align*}
G_n(\ES,\bJ):= & G(\ES,\mat{0}) + [\partial_{\bJ}G(\ES,\mat{0})](\bJ) + \frac{1}{2}[\partial^2_{\bJ}G(\ES,\mat{0})](\bJ,\bJ)\\
 & + \dots + \frac{1}{n!}[\partial^n_{\bJ}G(\ES,\mat{0})](\bJ,\dots,\bJ),
\end{align*}
where $\partial^n_{\bJ}G(\ES,\mat{0})$ is the $n$-th derivative of $G$ wrt $\bJ$ (a symmetric tensor of order $n$) evaluated at point $(\ES,\mat{0})$. All these terms can be evaluated, albeit laboriously. First, the fundamental relation $\partial_{J_{ij}}\ln Z_I = m_im_j + \partial^2_{\theta_i,\theta_j}\ln Z_I$ allows to replace derivatives wrt. $\bJ$ by derivatives wrt. $\btheta$. Second, 
at $\bJ=\mat{0}$, the Ising distribution boils down to a Bernoulli distribution, where all derivatives wrt. $\btheta$ are fully tractable.

The approximation at order $n=4$ is a classic computation \cite{VasilevRadzhabov1974,GeorgesYedidia1991,Nakanishi1997,Tanaka1998}, which yields~:
\begin{align}
G_4(\ES,\bJ)=& \sum_i G(m_i) - \sum_{(ij)}J_{ij}m_im_j 
- \frac{1}{2}\sum_{(ij)}J_{ij}^2c_ic_j \nonumber\\
&- \sum_{(ijk)}J_{ij}J_{jk}J_{ki}c_ic_jc_k - \frac{2}{3} \sum_{(ij)}J_{ij}^3m_ic_im_jc_j \nonumber\\
&- \sum_{(ijkl)}J_{ij}J_{jk}J_{kl}J_{li}c_ic_jc_kc_l\nonumber\\
&- 2 \sum_{(ij),k}J_{ij}^2J_{jk}J_{ki}m_ic_im_jc_jc_k \nonumber\\
&+ \frac{1}{12} \sum_{(ij)}J_{ij}^4c_ic_j(1+3m_i^2+3m_j^2-15m_i^2m_j^2).\label{eq:ising4}
\end{align}
Here, $(ij)$, $(ijk)$, $(ijkl)$ indicate respectively all unordered pairs, triplets and quadruplets of distinct spins. $G(m_i)$ is the free energy of each 1-spin marginal distribution (\refeq{Gmf} with $\theta_i=\tanh^{-1}(m_i)$). Finally, we use the shorthand $c_i=1-m_i^2$.

By differentiating this function wrt. $\ES$, we obtain a fixed point characterization of its extremum(s) $\ES$.
Stopping at order 1 in $\bJ$ yields the ``naive'' mean field equation. Stopping at order 2 (first line) yields the TAP equation. Stopping at order 3 (two first lines) yields \refeq{ising3} from the main text.


\subsubsection*{Bethe approximation}
In one particular case, $G(\ES,\bJ)$ in \refeq{Gmf} is tractable exactly. This is when the underlying couplings $J_{ij}$ define a \emph{tree topology}, that is, they are zero except on a subset of the edges defining a graph without loops.
In that case, the test Ising distribution $Q(\ES,\bJ)$ can be written as
\begin{equation}
\label{eq:Qtree}
Q(\bs|\ES,\bJ) = \prod_{\langle ij \rangle} \frac{Q(s_i,s_j)}{Q(s_i)Q(s_j)} \prod_i Q(s_i),
\end{equation}
where $\langle ij \rangle$ denotes all edges in the tree. This can be proved by repeated applications of Bayes' formula, starting from any leaf of the tree. 
Besides, each marginal $Q(s_i,s_j)$ is a 2-spin Ising distribution with coupling parameter $J_{ij}$, as proved directly by integrating out the remaining variables from the original Ising formula.

In consequence, the (exact) free energy writes
\begin{equation}
\label{eq:Gbethe}
G_{\rm B}(\ES,\bJ)=\sum_{i < j}G(m_i,m_j,J_{ij}) - (N-2)\sum_i G(m_i),
\end{equation}
$G(m_i)$ and $G(m_i,m_j,J_{ij})$ being the respective free energies of the marginal distributions $Q(s_i)$ and $Q(s_i,s_j)$, as defined by \refeq{Gmf}.
Note that the sum can be made over all spin pairs $i<j$, and not just neighboring pairs $\langle ij \rangle$ in the tree. Indeed, unconnected spin pairs yield a zero contribution, as $G(m_i,m_j,0)=G(m_i)+G(m_j)$.

The \emph{Bethe approximation} consists in using \refeq{Gbethe} as an approximation for the free energy, even when the $J_{ij}$ do not have a tree topology.

Imposing that $\partial_{\ES} G_{\rm B}=0$ and using \refeq{Gmf_m}, leads to the $N$ equations
\begin{equation}
\label{eq:bethe_statio}
\forall i, \quad (N-2)\tanh^{-1}(m_i) = \big(\sum_j \tcavij\big) -h_i,
\end{equation}
where  the so-called \emph{cavity fields} $(\tcavij,\tcavji)$ are the natural field parameters of each 2-spin Ising distribution $Q(s_i,s_j)$, that is, tractable functions of $(m_i,m_j,J_{ij})$.
In fact, it is easily shown that the 2-spin Ising distribution of natural parameters $(\tcavij,\tcavji,J_{ij})$ has moments
\begin{equation}
\label{eq:2spin_moments}
m_i = \frac{t_i+t_jt_{ij}}{1+t_it_jt_{ij}} \quad,\quad
c_{ij} = t_{ij}\frac{(1-t_i^2)(1-t_j^2)}{(1+t_{ij}t_it_j)^2},
\end{equation}
with $t_{ij}=\tanh(J_{ij})$, $t_i=\tanh(\tcavij)$, $t_j=\tanh(\tcavji)$. Inserting this expression for $m_i$ into \refeq{bethe_statio}, with $\tanh^{-1}(m)=\frac{1}{2}(\ln(1+m)-\ln(1-m))$, and after some linear recombinations, we obtain
\begin{equation}
\label{eq:FPBP}
\forall (i,j),\quad \tcavij = h_i + \sum_{k\neq j}\tanh^{-1}\big(t_{ik}\tanh(\theta_k^{(i)})\big),
\end{equation}
and the optimum is now characterized by $N(N-1)$ equations over the $N(N-1)$ cavity variables.
This switching from $N$ principal variables (the $m_i$) to $N(N-1)$ auxiliary variables (the $\tcavij$) can also be interpreted as a dual Lagrangian optimization procedure \cite{Yedidia2001}.

In a tree-like topology, \refeq{FPBP} can be solved iteratively starting from any leaf of the tree, allowing to recover the exact values for all the $\tcavij$, and thus, for magnetizations $m_i^\star$. The resulting algorithm is known as \emph{belief propagation}, or \emph{sum-product}.
In a general topology, the fixed point approach to characterize solutions of \refeq{FPBP} is known as \emph{loopy belief propagation}. It is not guaranteed to have a single solution anymore -- and it only characterizes an approximation  for the magnetizations $m_i$.

The covariances $C_{ij}$ can be approximated in turn, based on the linear response formula, \refeq{linresp}. Applied to each 2-spin distribution of natural parameters $(\tcavij,\tcavji,J_{ij})$, it implies the differential equality
\begin{equation}
\label{eq:2spinlinresp}
\forall (i,j),\quad (\partial m_i) = (1-m_i^2)\big(\partial \tcavij\big) + c_{ij}\big(\partial \tcavji\big)
\end{equation}
with $c_{ij}$ given by \refeq{2spin_moments}.
We can then differentiate \refeq{bethe_statio} as a function of $(\partial h_i)$ and linearly eliminate the cavity fields $(\partial \tcavij)$ thanks to \refeq{2spinlinresp}. As a result, we express $(\partial h_i)$ as a function of the $(\partial m_j)$ only, and this provides the linear response prediction~:
\begin{align}
(\CovS^{-1}_{\rm B})_{ij} =& \left(1+\sum_k \frac{c_{ik}^2}{(1-m_i^2)(1-m_k^2)-c_{ik}^2}\right)\frac{\delta_{ij}}{1-m_i^2} \nonumber\\& - \frac{c_{ij}}{(1-m_i^2)(1-m_j^2)-c_{ij}^2}.\label{eq:bethe_linresp}
\end{align}
This derivation, which I could not find in the literature, expresses the linear response matrix $\CovS_{\rm B}$ in an alternative 
form than in \cite{Ricci2012}.

To derive the weak coupling (Plefka) expansion of the Bethe approximation, note that each $G(m_i,m_j,J_{ij})$ in \refeq{Gbethe} is an exact Ising free energy over two spins $i$ and $j$. Compared to a generic Ising free energy over $N$ spins, its expansion only contains the two-spin diagrams, summed over the single spin pair involved.
It follows, after summing over all spin pairs $i<j$ in \refeq{Gbethe}, that the Plefka expansion of the Bethe free energy is obtained by keeping only the pairwise diagrams in the expansion for the true Ising free energy.





\bigskip

\section{Cavity method and adaptive TAP equations}
\label{app:adaptiveTAP}

\subsubsection*{Cavity method}

The \emph{cavity method} is a classic approach allowing to recover many analytical properties of the Ising model, and other multivariate exponential models \cite{Mezard1987,OpperSaad2001}. Singling out an arbitrary spin location $i$, one can rewrite \refeq{ising1} as
\begin{equation*}
P(\bs\cavi,s_i)\sim P\cavi(\bs\cavi) \exp\Big(s_i\big(h_i+\sum_{j \neq i} J_{ij}s_j\big)\Big),
\end{equation*}
where $\bs\cavi$ denotes the remaining $N-1$ spins, and the so-called \emph{cavity distribution} $P\cavi$ is the Ising distribution obtained by deleting line and column $i$ from $(\bh,\bJ)$. The remaining spins interact with $i$ only through the random variable
\begin{equation}
\label{eq:hcav}
\hcav_i := h_i + \sum_{j\neq i}J_{ij}s_j,
\end{equation}
and we can write
\begin{equation}
\label{eq:cavity}
P(\hcav_i,s_i)\sim P\cavi(\hcav_i) \exp(s_i \hcav_i),
\end{equation}
$P\cavi(\hcav_i)$ indicating the distribution of variable $\hcav_i$ when the spins $\bs\cavi$ follow the cavity distribution $P\cavi$.

In general, distribution $P\cavi(\hcav_i)$ is not tractable exactly \footnote{Except when the couplings $J_{ij}$ have a tree-like topology. In this case, $P\cavi(\hcav_i)$ is a factorized product over the neighboring spins of $i$, and this is another way of deriving the Bethe equation \refno{FPBP} \cite{MezardParisi2001}.}.
But in many circumstances, since $\hcav_i$ is the sum of variables with many degrees of freedom, it can be assumed to have a normal distribution~:
\[
P\cavi(\hcav_i) = \Norm(\hcav_i|\theta_i,V_i),
\]
and \refeq{cavity} becomes, approximately~:
\begin{equation}
\label{eq:cavityapp}
P(\hcav_i,s_i)\sim \exp\left(-\frac{1}{2V_i}(\hcav_i-\theta_i)^2 + s_i \hcav_i\right).
\end{equation}

This equation is the starting point of the classic cavity method. Note the formal similarity with our definition for the joint probability of spins and latent fields, \refeq{IL}, so we can simply recycle our results. From \refeq{ising}, $P(s_i)$ is the Bernoulli distribution of parameter $\theta_i$, so $m_i=\tanh(\theta_i)$. From \refeq{ERS}, we have
\(
\pE(\hcav_i) = \theta_i + V_im_i.
\)
Taken together, this yields the generalized TAP equation~: 
\begin{equation}
\label{eq:TAPgen}
m_i = \tanh\big(h_i+\sum_j J_{ij}m_j - m_i V_i \big).
\end{equation}

For the SK model in the limit $N\rightarrow\infty$, it can be shown that the different summands to $\hcav_i$ in the cavity distribution (\refeq{hcav}) become linearly independent \cite{Mezard1987,Nishimori2001}, so the variance writes $V_i \simeq \sum_j J_{ij}^2(1-m_j^2)$ and we recover the classic TAP equation.

\subsubsection*{Adaptive TAP approximation}

Instead, in the \emph{adaptive TAP} method \cite{OpperWinther2001,OpperWinther2001b}, $V_i$ is left as a free variable which can adapt to any statistical structure of the couplings $\bJ$. First, differentiating \refeq{TAPgen} wrt. $\bh$ (but neglecting the dependency of $V_i$ itself) leads to a linear response prediction for $\CovS$~:
\begin{align}
\label{eq:linrespgen}
(\CovS^{-1}_{\rm A})_{ij} &= \big(1+(1-m_i^2)V_i\big)\frac{\delta_{ij}}{1-m_i^2}- J_{ij}.
\end{align}
Second, given magnetization $m_i$, the individual variance of spin $i$ should be $1-m_i^2$. Self-coherence of the variance prediction imposes that
\begin{align}
\label{eq:FPgen}
1-m_i^2 &= (\CovS_{\rm A})_{ii}.
\end{align}
Taken together, \refeq{TAPgen}-\refno{FPgen} constitute a system on variables $\{m_i,V_i\}$, which can be solved by classic iterative methods \cite{OpperWinther2001b}.

\medskip

I now turn to the weak coupling (Plefka) expansion of \refeq{TAPgen}-\refno{FPgen}, when magnetizations $m_i$ are fixed, and the coupling matrix writes $\alpha\bJ$.
Given the form of \refeq{TAPgen}, this only requires to obtain the expansion for $m_iV_i$ or, after a convenient rescaling, for variable
\[
x_i := V_i(1-m_i^2),
\]
for which we want to establish the Taylor development
\[
x_i = \alpha x_i^{[1]} + \alpha^2 x_i^{[2]} + \alpha^3 x_i^{[3]} + \dots
\]
(note that $x_i=V_i=0$ when $\alpha=0$).

From \refeq{linrespgen}, it is clear that the solution depends on the diagonal of $\bJ$ only through the simple offset $V_i\rightarrow V_i+J_{ii}$, so we may assume $J_{ii}=0$ without loss of generality.

Introducing the variables
\begin{align*}
d_i &:= \frac{1-m_i^2}{1+(1-m_i^2)V_i}= (1-m_i^2)\big[1-x_i+x_i^2-x_i^3+\dots\big],
\end{align*}
and matrix $\mD=\diag(d_i)$, we rewrite \refeq{linrespgen} as
\[
\CovS^{-1} = \mD^{-1}-(\alpha\bJ)
\]
and thus, after a classic switching from $\CovS^{-1}$ to $\CovS$~:
\begin{align*}
\CovS 
&= \mD + \alpha \mD\bJ\mD + \alpha^2 \mD(\bJ\mD)^2 + \alpha^3 \mD(\bJ\mD)^3 + \dots,
\end{align*}
allowing to easily express the development of $\CovS$ from that of $d_i$.

Then, noting $c_i:=1-m_i^2$ for concision, the fixed point equation \refno{FPgen} imposes, at order 4~:
\begin{align*}
1 &= \Big(1-x_i+x_i^2-x_i^3+x_i^4\Big)\\
&+ \alpha^2 \sum_j c_iJ_{ij}^2 c_j \Big(1-2x_i+3x_i^2-x_j+2x_ix_j+x_j^2\Big)\\
&+ \alpha^3 \sum_{j,k} c_iJ_{ij}c_j J_{jk}c_kJ_{ki} \Big(1-2x_i-x_j-x_k\Big)\\
&+ \alpha^4 \sum_{j,k,l} c_iJ_{ij}c_j J_{jk}c_kJ_{kl}c_lJ_{li} \quad + o(\alpha^4).
\end{align*}
We can then replace $x_i$ by its expansion, and regroup the powers of $\alpha$. For the equation to be verified at order 1, this imposes that
\[
x_i^{[1]} = 0.
\]
Using this newly found value, the fixed point equation at order 2 imposes
\[
\frac{m_i}{c_i}x_i^{[2]} = m_i\sum_{j\neq i} J_{ij}^2c_j.
\]
Then, the fixed point equation at order 3 imposes
\[
\frac{m_i}{c_i}x_i^{[3]} = 2m_i \sum_{(jk|i)} J_{ij}J_{jk}J_{ki}c_jc_k.
\]
where $(jk|i)$ denotes all unordered triplets of the form $\{i,j,k\}$ with $j$ and $k$ distinct, and distinct from $i$.
By inserting these values into \refeq{TAPgen}, we recover the expansion from the main text, \refeq{adap3}.

Finally, the fixed point equation at order 4 yields
\begin{equation}
\label{eq:adap4}
\frac{m_i}{c_i}x_i^{[4]} = 2 m_i\sum_{(jkl|i)} J_{ij}J_{jk}J_{kl}J_{li}c_jc_kc_l - m_ic_i\sum_{j\neq i}J_{ij}^4c_j^2.
\end{equation}
Note that the sum over $(jkl|i)$ -- which involves the most terms and is generally dominant -- is identical to that for the true Ising expansion~: see \refeq{ising4}.

\bigskip

\section{Weak coupling expansion for the Cox approximation}
\label{app:perturb}


We consider the `variational' Cox approximation, solution to the equations
\begin{align}
m_i &= \int_{x\in\er} \tanh\left(\mu_i+x\sqrt{\Sigma_{ii}}\right)\phi(x)\dx, \label{eq:EScox_}\\
d_i &= \int_{x\in\er} \left(1-\tanh^2\left(\mu_i+x\sqrt{\Sigma_{ii}}\right)\right)\phi(x)\dx, \label{eq:di_}\\
(\CovS^{-1})_{ij} &= d_i^{-1}\delta_{ij} -\alpha J_{ij}.\label{eq:CovS_var_}\\
\ER &= \bh+\alpha\bJ\ES, \label{eq:ERS_} \\
\CovR  &= \alpha\bJ+\alpha^2\bJ\CovS\bJ, \label{eq:CovRS_}
\end{align}
when magnetizations $m_i$ are fixed, and the coupling matrix writes $\alpha\bJ$, $\alpha$ being the small parameter of the expansion.

Here, I detail the computation up to order 3, and also provide the result at order 4. The overall structure of the computation is largely similar to that for the adaptive TAP approximation, in the previous paragraph.



When $\alpha=0$, the solution is obvious~: couplings $\alpha\bJ$ vanish, and so does the covariance matrix $\CovR$. The Cox distribution $Q(\ER,\CovR)$ is simply a Bernoulli distribution $\Bern(\ER)$ with $\ER=\bh$, and the fixed point equations impose that
\begin{align*}
\mu_i &= \tanh^{-1}(m_i),\\
d_i &= 1-m_i^2.
\end{align*}

We now seek a Taylor expansion for the solution of \refeq{EScox_}-\refno{CovRS_} when $\alpha$ is small but nonzero,  and magnetizations $m_i$ are fixed.
More precisely, noting
\[
\Delta_i:=\mu_i-\tanh^{-1}(m_i),
\]
our purpose is to find the parameters in the following Taylor expansions~:
\begin{align}
\Delta_i &=  \alpha\Delta_i^{[1]} + \alpha^2\Delta_i^{[2]} + \alpha^3\Delta_i^{[3]} + \dots \nonumber\\ 
\Sigma_{ij} &= \alpha\Sigma_{ij}^{[1]} + \alpha^2\Sigma_{ij}^{[2]} + \alpha^3\Sigma_{ij}^{[3]} + \dots \nonumber
\end{align}
When inserted into \refeq{ERS_}, the development of $\Delta_i$ will exactly provide the desired Plefka expansion.

\subsubsection*{Development for equation \refno{EScox_}}

Equation \refno{EScox_} writes
\[
m_i = \int_{x\in\er} \tanh\left(\tanh^{-1}(m_i)+\Delta_i+x\sqrt{\Sigma_{ii}}\right)\phi(x)\dx,
\]
where $\Delta_i$ is of leading order $\alpha$, and $\sqrt{\Sigma_{ii}}$ is of leading order $\alpha^{1/2}$.
Applying the Taylor development of $\tanh$~:
\begin{align*}
\tanh\left(\tanh^{-1}(m)+X\right) =& m + (1-m^2)\Big[X-mX^2+\dots \Big]
\end{align*}
up to order 6 (because $\sqrt{\Sigma_{ii}}$ is of leading order $\alpha^{1/2}$), and using the classic integration formulas~:
\begin{align}
\label{eq:phimoments}
\int_x x^n\phi(x)\dx = \begin{cases}
(n-1)(n-3)\dots & \textrm{if $n$ is even} \\
0 & \textrm{if $n$ is odd}
\end{cases}
\end{align}
we obtain
\begin{align*}
m_i =& m_i + (1-m_i^2)\Big[\Delta_i 
-m_i \big(\Delta_i^2+\Sigma_{ii}\big) \\
& + \big(m_i^2-\tfrac{1}{3}\big)\big(\Delta_i^3+3\Delta_i\Sigma_{ii}\big) \\
& + \big(-m_i^3+\tfrac{2}{3}m_i\big)\big(\Delta_i^4 + 6\Delta_i^2\Sigma_{ii} + 3\Sigma_{ii}^2\big)\\
& + \big(m_i^4-m_i^2+\tfrac{2}{15}\big)\big(10 \Delta_i^3\Sigma_{ii} + 15 \Delta_i\Sigma_{ii}^2\big) \\
& + \big(-m_i^5+\tfrac{4}{3}m_i^3-\tfrac{17}{45}m_i\big) \big(45 \Delta_i^2 \Sigma_{ii}^2 + 15 \Sigma_{ii}^3\big)\Big]+o(\alpha^3).
\end{align*}
Notice that, after integration by the Gaussian kernel, only integer powers of $\Sigma_{ii}$ remain.

\begin{widetext}
For this equation to be verified, the term inside square brackets must be equal to zero up to order $\alpha^3$. Expanding $\Delta_i$ and $\Sigma_{ii}$ with the shorthand $X_k=\Delta_i^{[k]}$, $Y_k=\Sigma_{ii}^{[k]}$, and regrouping the powers of $\alpha$, we obtain~:
\begin{align}
0 &= \alpha\Big[X_1 - m_i Y_1\Big] \label{eq:ord1}\\
& + \alpha^2\Big[X_2 - m_i (X_1^2+Y_2) + (3m_i^2-1)X_1Y_1 + (-3m_i^3+2m_i)Y_1^2\Big] \label{eq:ord2}\\
& + \alpha^3\Big[X_3 - m_i (2X_2X_1+Y_3) + (m_i^2-\tfrac{1}{3})(X_1^3+ 3X_2Y_1 + 3X_1Y_2) + (-6m_i^2+4m_i)(X_1^2Y_1+Y_2Y_1)
\nonumber\\ & \qquad 
+ (15m_i^4-15m_i^2+2)X_1Y_1^2 + (-15m_i^5+20m_i^3-\tfrac{17}{3}m_i)Y_1^3 \Big] \label{eq:ord3}
&+ o(\alpha^3).
\end{align}
\end{widetext}

\subsubsection*{Solution at order 2}

At this point, we can readily find the two first orders of the solution. Indeed, we have $\CovR=\alpha\bJ+\alpha^2\bJ\CovS\bJ$ and $C_{ij}=(1-m_i^2)\delta_{ij}+o(1)$, and so
\begin{align*}
\Sigma_{ij}^{[1]} &= J_{ij}, \\
\Sigma_{ij}^{[2]} &= \sum_k J_{ik}J_{jk}(1-m_k^2).
\end{align*}

Then, at order 1, the fixed point equation above (line \refno{ord1}) imposes that
\[
\Delta_i^{[1]} = m_i \Sigma_{ii}^{[1]} \quad = m_iJ_{ii}.
\]

Using this new value, the fixed point equation at order 2 (line \refno{ord2}) imposes that
\begin{align*}
\Delta_i^{[2]} &= m_i \big((\Delta_i^{[1]})^2 + \Sigma_{ii}^{[2]}\big) - (3m_i^2-1)\Delta_i^{[1]}\Sigma_{ii}^{[1]} +\dots\\ 
& = m_i \sum_{j\neq i}J_{ij}^2(1-m_j^2),
\end{align*}
which is identical to the order 2 coefficient in the exact Ising model -- see \refeq{ising3}. Note that the diagonal terms $J_{ii}$ are nonzero and an active part of the derivation, but cancel out in the final result, so they play no role in the expansion up to order 2.

\subsubsection*{Development for $\CovS$}

In general, to establish the development of $\CovR$ at any given order $n$, we need the expansion of $\CovS$ up to order $n-2$, because $\CovR=\alpha \bJ+\alpha^2\bJ\CovS\bJ$. Thus, to expand $\ER$ and $\CovR$ at order 3, we must first establish the development for $\CovS$ at order 1,
based on the development  of $(\ER,\CovR)$ at order 1 established above.

Regarding \refeq{di_}, a very similar computation (development of $\tanh$ at order 2, and simplification by the integral formulas of \refeq{phimoments}) yields~:
\begin{align*}
d_i &= (1-m_i^2) - \alpha(1-m_i^2)^2J_{ii} +o(\alpha).
\end{align*}
Introducing matrix $\mD={\rm diag}(d_i)$, \refeq{CovS_var_} writes
\[
\CovS^{-1} = \mD^{-1}-(\alpha\bJ),
\]
and thus, after a classic switching from $\CovS^{-1}$ to $\CovS$~:
\begin{align*}
\CovS 
&= \mD + \alpha \mD\bJ\mD + o(\alpha).
\end{align*}
So finally~:
\begin{align}
C_{ij} &= \delta_{ij}(1-m_i^2) \nonumber\\
& +  \alpha (1-\delta_{ij})(1-m_i^2)(1-m_j^2) J_{ij} + o(\alpha). \label{eq:C_ord1}
\end{align}

\subsubsection*{Solution at orders 3 and 4}

The coefficient $C_{ij}^{[1]}$ found in \refeq{C_ord1} 
is pasted into $\CovR=\alpha\bJ+\alpha^2\bJ\CovS\bJ$, to obtain~:
\begin{equation}
\Sigma_{ij}^{[3]} = \sum_{k\neq l} J_{ik}J_{kl}J_{lj}(1-m_k^2)(1-m_l^2).
\end{equation}
Then, line \refno{ord3} allows to find the order 3 coefficient of $\ER$. 
After computation, this gives~:
\begin{align*}
\Delta_i^{[3]} &= 2m_i\sum_{(jk|i)}J_{ij}J_{jk}J_{ki}(1-m_j^2)(1-m_k^2)\\
& - 2(m_i^2-\tfrac{1}{3})J_{ii}^3m_i(1-m_i^2),
\end{align*}
where $(jk|i)$ denotes all unordered triplets of the form $\{i,j,k\}$ with $j$ and $k$ distinct, and distinct from $i$.

Using the values found for $\Delta_i^{[1]}$,  $\Delta_i^{[2]}$ and  $\Delta_i^{[3]}$, and the fact that $\mu_i = h_i + \alpha\sum_j J_{ij}m_j$, yields \refeq{cox3} from the main text.

Pushing all computations one order further, with the help of the computer algebra system MAXIMA, yields~:
\begin{align}
\label{eq:cox4}
\Delta_i^{[4]} &= 2m_i \sum_{(jkl|i)} J_{ij} J_{jk} J_{kl} J_{li} c_jc_kc_l 
-m_ic_i \sum_{j\neq i}J_{ij}^4c_j^2  \nonumber\\
& -2m_i c_i(3m_i^2-1)J_{ii}^2 \sum_{j\neq i} J_{ij}^2 c_j  \nonumber\\
& -2m_i c_i(7m_i^4-8m_i^2+\tfrac{5}{3})J_{ii}^4  \nonumber\\
& -2m_i \sum_{j\neq i} J_{ij}^2 J_{jj}^2 m_j^2c_j^2,
\end{align}
with the shorthand $c_i=1-m_i^2$. The two first terms are identical to the adaptive TAP expansion, \refeq{adap4}.
The remaining terms involve the diagonal weights $J_{ii}$, and would be absent if the coupling matrix was such that ${\rm diag}(\bJ)=\mat{0}$.

\bigskip

\section{Approximate formulas for the Cox distribution}
\label{sec:tanherf}

The formulas inherent to the Cox distribution, \refeq{KLQP}, \refno{ES_var}, \refno{di} and \refno{CovScox} from the main text, are easily estimated by numerical integration (for example, Simpson quadrature). But the overall computation time quickly becomes forbidding, as these estimations must be done for each pair of spins, and on many iterations to target the fixed point.

Hence, I found it more convenient to use approximate formulas. Let us note $L$ the logistic function at scale $1/2$, that is~:
\[
L(r):=\frac{1}{1+e^{-2r}}.
\]
Function $L$ is pivotal in the Bernoulli distribution, since $\tanh(r)=2L(r)-1$, $\log2\cosh(r)=2\int_0^r L(u){\rm d}u-r$, and $1-\tanh^2(r)=2L'(r)$.

I suggest to approximate $L$ by the following combination of Gaussian functions~:
\begin{equation}
\label{eq:Lapp}
L^{\rm app}(r) = \Phi\left(\frac{r}{\tkappa}\right) + \tnu \phi'\left(\frac{r}{\tlambda}\right),
\end{equation}
with $\Phi(x):=\int_{-\infty}^x\phi(u){\rm d}u$ the standard normal cumulative distribution, and $\phi'(r)=-r\phi(r)$.

Taking parameters $(\tkappa,\tnu,\tlambda)\simeq(0.7072 , 0.1648 , 0.9712)$, one has $\|L -L^{\rm app}\|_\infty<0.001$ on the whole real line. The approximation also applies to the primitive, with $\|\int (L - L^{\rm app})\|_\infty<0.001$, and to the first derivative, with $\|L' - (L^{\rm app})'\|_\infty<0.0033$.

As the convolution product of two Gaussian functions remains Gaussian, this replacement allows to compute analytically all the formulas. Here, I only provide the results, and refer to Supplementary Material for the derivation.

Given spin index $i$, let us introduce the following reduced quantities~:
\begin{align*}
x_i &:= \frac{\mu_i}{\sqrt{\Sigma_{ii}}},\\
k_i &:= \frac{\sqrt{\Sigma_{ii}}}{\sqrt{\tkappa^2+\Sigma_{ii}}}, \\
l_i &:= \frac{\sqrt{\Sigma_{ii}}}{\sqrt{\tlambda^2+\Sigma_{ii}}}.
\end{align*}

Then, the first moment of the Cox distribution, \refeq{ES_var} (or equivalently \refeq{EScox}), can be approximated as
\begin{equation}
\label{eq:miapp}
m_i^{\rm app} = 2 \left[\Phi\left(x_ik_i\right) + \tnu (1-l_i^2)\phi'\left(x_il_i\right)\right] - 1
\end{equation}
with a guaranteed maximum error $|m_i-m_i^{\rm app}|<0.002$.

The variance term $d_i$ in \refeq{di} is approximated as
\begin{equation*}
d_i^{\rm app} = 2\Sigma_{ii}^{-1/2} \left[k_i\phi(x_ik_i) + \tnu (1-l_i^2)l_i\phi''(x_il_i)\right]
\end{equation*}
with a guaranteed maximum error $|d_i-d_i^{\rm app}|<0.007$.

To concretely estimate the free energy associated to \refeq{KLQP}, it is necessary to compute $F_i:=\int_x \log2\cosh(\mu_i+x\sqrt{\Sigma_{ii}})\phi(x)\dx$. It is approximated as
\begin{equation*}
F_i^{\rm app}=\mu_i\left[2\Phi(x_ik_i)-1\right] + 2\sqrt{\Sigma_{ii}} \left[\frac{\phi(x_ik_i)}{k_i} - \tnu  \frac{\phi''(x_il_i)}{l_i}\right]
\end{equation*}
with a guaranteed maximum error inferior to $0.002$.

The approximate formula for the covariance of the Cox distribution, \refeq{CovScox}, is quite bulky and  provided in Supplementary Material. It has guaranteed maximum error inferior to $0.008$.

In my numerical tests, using these approximate formulas instead of lengthier Simpson quadrature yielded no noticeable difference in the final solution of the fixed point equations. At the same time, computation times were cut by (up to) two orders of magnitude.




%

\end{document}